\documentclass[aps,prd,twocolumn,superscriptaddress,showpacs,preprintnumbers,draft]{revtex4}

\usepackage{array}
\usepackage{dcolumn}
\usepackage{graphics}

\input boxedeps
\SetepsfEPSFSpecial
\HideDisplacementBoxes

\newcommand{\slj}[3]{\mbox{$^{#1}${\ifcase#2\or S\or 
         P\or D\or F\or G\fi}$_{#3}$}}
\newcommand{\sLj}[3]{{}^{#1}\!#2_{#3}}
	 
\newcommand{\kev}{\hbox{ keV}}
\newcommand{\mev}{\hbox{ MeV}}
\newcommand{\gev}{\hbox{ GeV}}

\newcommand{\zcc}{\mathcal{Z}_{c\bar{c}}}

\newcommand{\cfrac}[2]{\textstyle{\frac{#1}{#2}}}

\newcommand{\jpsi}{\ensuremath{J\!/\!\psi}}

\def\bra#1{\langle #1|}
\def\ket#1{\,| #1\rangle}

\begin{document}

\preprint{FERMILAB--Pub--04/001--T}
\preprint{BUHEP-04-01}

\title{Charmonium levels near threshold and the narrow state $X(3872) \to \pi^{+}\pi^{-}\jpsi$}


\author{Estia J. Eichten}
\email[E-mail: ]{eichten@fnal.gov}
\affiliation{Theoretical Physics Department\\ Fermi National 
Accelerator Laboratory\\ P.O.\ Box 500, Batavia, IL 60510}
\author{Kenneth Lane}
\email[E-mail: ]{lane@bu.edu}
\affiliation{Department of Physics, Boston
University\\ 590 Commonwealth Avenue, Boston, MA 02215}
\author{Chris Quigg}
\email[E-mail: ]{quigg@fnal.gov}
\affiliation{Theoretical Physics Department\\ Fermi National 
Accelerator Laboratory\\ P.O.\ Box 500, Batavia, IL 60510}


\date{\today}

\begin{abstract}
We explore the influence of open-charm channels on charmonium properties, and 
profile the 1\slj{3}{3}{2}, 1\slj{3}{3}{3}, and 
2\slj{1}{2}{1} charmonium candidates for $X(3872)$. The favored 
candidates, the 1\slj{3}{3}{2} and 1\slj{3}{3}{3} levels, both have 
prominent radiative decays. The 1\slj{3}{3}{2} might be visible in 
the $D^{0}\bar{D}^{*0}$ channel, while the dominant decay of the 
1\slj{3}{3}{3} state should be into $D\bar{D}$. We propose that
additional discrete charmonium levels can be discovered as narrow resonances of 
charmed and anticharmed mesons.
\end{abstract}


\maketitle

\section{Introduction\label{sec:intro}}
%
Encouraged by the Belle Collaboration's sighting~\cite{Choi:2002na} of
$\eta_{c}^{\prime}(2\slj{1}{1}{0})$ in exclusive $B \to K K_{S} K^{-}\pi^{+}$
decays, we  sketched a coherent strategy  to explore 
$\eta_{c}^{\prime}$ and the remaining  charmonium states that do not 
decay into open charm,
$h_{c}(1\slj{1}{2}{1})$,
$\eta_{c2}(1\slj{1}{3}{2})$, and
$\psi_{2}(1\slj{3}{3}{2})$, through $B$-meson
gateways~\cite{Eichten:2002qv}.  We argued that
 radiative transitions among charmonium levels and $\pi\pi$
cascades to lower-lying charmonia would enable the
identification of these states.

Now the Belle Collaboration
has presented evidence~\cite{Choi:2003ue} for a new narrow state,
$X(3872) \to \pi^{+}\pi^{-}\jpsi$, seen in $B^{\pm} \to
K^{\pm}X(3872)$.  The CDF
Collaboration has confirmed the new state in
inclusive 1.96-TeV $\bar{p}p \to 
X(3872)+\hbox{anything}$~\cite{Acosta:2003zx}, as has the D\O\ 
experiment~\cite{Dzerotemp}.  
In addition, the
CLEO~\cite{Asner:2003wv}, BaBar~\cite{Wagner:2003qb}, and 
Belle~\cite{Abe:2003ja} 
experiments have confirmed and refined the discovery of
$\eta_{c}^{\prime}$, fixing its mass and width as $M(\eta_{c}^{\prime})
= 3637.7 \pm 4.4 \mev$ and $\Gamma(\eta_{c}^{\prime}) = 19 \pm
10\mev$~\cite{Skwarnicki:2003wn}.  

In this Article, we develop the hypothesis that $X(3872)$ is a
charmonium level.  The new meson's position at $D^{0}\bar{D}^{*0}$
threshold makes it imperative to take account of the coupling between
$c\bar{c}$ bound states and open-charm channels.  Accordingly, we
revisit the properties of charmonium levels, using the Cornell
coupled-channel model~\cite{Eichten:1978tg,Eichten:1980ms} to assess
departures from the single-channel potential-model expectations.  Far
below charm threshold, the nonrelativistic potential model is a good
approximation to the dynamics of the charm-anticharm system.  For
excited states above the first few levels, the coupling of $c\bar{c}$
to charmed-meson pairs modifies wave functions, masses, and transition
rates.  We estimate spin-splittings induced by communication with
open-charm channels, and examine the effect of configuration mixing on
radiative decay rates.  We consider $\psi_{2}$ (1\slj{3}{3}{2}),
$\psi_{3}$ (1\slj{3}{3}{3}), and $h_{c}^{\prime}$ (2\slj{1}{2}{1}) as
possible interpretations of $X(3872)$, commenting briefly on
diagnostics of a general character that will help establish the nature
of $X(3872)$.  Independent of the identity of $X(3872)$,
above-threshold charmonium states should be visible as narrow
structures in $1\slj{3}{3}{3} \to D\bar{D}$, $2\slj{3}{2}{2} \to
D\bar{D}, D\bar{D}^{*}$, $1\slj{3}{4}{4} \to D\bar{D},
D\bar{D}^{*}$, and possibly $2\slj{3}{2}{0} \to D\bar{D}$.

\textit{What do we know about $X(3872)$?} Belle's clean sample of 36 
events, entirely from $B$-meson decays, determines the mass 
of the new state as $3872.0 \pm 0.6 \pm 0.5\mev$, and yields a ratio of 
production $\times$ decay branching fractions, 
\begin{equation}
    \frac{\mathcal{B}(B^{+} \!\to \!K^{+}X) 
    \mathcal{B}(X \!\to \!
    \pi^{+}\pi^{-}\jpsi)}{\mathcal{B}(B^{+}\!\to\! K^{+}\psi^{\prime}) 
    \mathcal{B}(\psi^{\prime}\!\to\! \pi^{+}\pi^{-}\jpsi)} = 
    0.063 \pm 0.014\; .
\end{equation}
CDF observes $730 \pm 90$ events above background and determines a 
mass $3871.4 \pm 0.7 \pm 0.4\mev$. The observed mass lies $67\mev$ 
above the 1\slj{3}{3}{J} centroid in the potential-model template of 
Ref.~\cite{Eichten:2002qv}.
The large number of events suggests 
that much of the CDF sample arises from prompt production of 
$X(3872)$, not from $B$ decays, and opens another path to the 
exploration of new charmonium states.
Belle sets a 90\%\ C.L.\ upper limit on the width, $\Gamma(X(3872))<2.3\mev$.
The $\pi^{+}\pi^{-}\jpsi$ decay appears to favor high dipion masses,
but there is no detailed information yet about the quantum numbers
$J^{PC}$.  Belle has searched in vain for radiative transitions to
the 1\slj{3}{2}{1} level; their 90\%\ C.L.\ upper bound,
\begin{equation}
    \frac{\Gamma(X(3872) \to \gamma \chi_{c1})}{\Gamma(X(3872) \to 
    \pi^{+}\pi^{-}\jpsi)} < 0.89\; ,
    \label{eqn:radbound}
\end{equation}
conflicts with our single-channel potential-model expectations
 for the 1\slj{3}{3}{2} state~\cite{Eichten:2002qv}. The theoretical 
 estimate of the $\pi\pi\jpsi$ rate is highly uncertain, however.
 
 \section{Influence of open-charm states \label{sec:OC}} The Cornell group showed long ago that a very simple model that
 couples charmonium to charmed-meson decay channels confirms the
 adequacy of the single-channel $c\bar{c}$ analysis below threshold and
 gives a qualitative understanding of the structures observed above
 threshold~\cite{Eichten:1978tg,Eichten:1980ms}. We now employ the 
 Cornell coupled-channel formalism to analyze the properties of 
 charmonium levels that populate the threshold region between $2M(D)$ 
 and $2M(D^{*})$, for which the main landmarks are shown in 
 Table~\ref{table:thresholds}.
 \begin{table}
 \caption{Thresholds for decay into open charm.\label{table:thresholds}}
 \begin{ruledtabular}
     \begin{tabular}{c d}
 Channel & \multicolumn{1}{c}{Threshold Energy (MeV)} \\
 \hline
 & \\[-6pt]
 \multicolumn{1}{c}{$D^{0}\bar{D}^{0}$}  &  
 3729.4 \\
 \multicolumn{1}{c}{$D^{+}D^{-}$} & 
 3738.8  \\
 \multicolumn{1}{c}{$D^{0}\bar{D}^{*0} \hbox{ or } D^{*0}\bar{D}^{0}$} & 
   3871.5  \\
 \multicolumn{1}{c}{$D^{\pm}D^{*\mp}$} & 
 3879.5  \\
 \multicolumn{1}{c}{$D_{s}^{+}D_{s}^{-}$} &   3936.2 \\
 \multicolumn{1}{c}{$D^{*0}\bar{D}^{*0}$} &  4013.6 \\
 \multicolumn{1}{c}{$D^{*+}D^{*-}$} &  4020.2 \\
 \multicolumn{1}{c}{$D_{s}^{+}\bar{D}_{s}^{*-} \hbox{ or }
 D_{s}^{*+}\bar{D}_{s}^{-}$} &   4080.0  \\
 \multicolumn{1}{c}{$D_{s}^{*+}D_{s}^{*-}$} &   4223.8 \\
     \end{tabular}
 \end{ruledtabular}
 \end{table}

 Our  command of quantum chromodynamics is inadequate to derive
 a realistic description of the interactions that communicate between
 the $c\bar{c}$ and $c\bar{q}+\bar{c}q$ sectors.  The Cornell formalism
 generalizes the $c\bar{c}$ model 
 without introducing new parameters, writing the interaction 
 Hamiltonian in second-quantized form as
 \begin{equation}
     \mathcal{H}_{I} = \cfrac{3}{8} \sum_{a=1}^{8} 
     \int:\rho_{a}(\mathbf{r}) V(\mathbf{r} - 
     \mathbf{r}^{\prime})\rho_{a}(\mathbf{r}^{\prime}): 
     d^{3}{r}\,d^{3}{r}^{\prime}\; ,
     \label{eq:CCCMH}
 \end{equation}
 where $V$ is the charmonium potential and $\rho_{a}(\mathbf{r}) = 
 \cfrac{1}{2}\psi^{\dagger}(\mathbf{r})\lambda_{a}\psi(\mathbf{r})$ is the color 
 current density, with $\psi$ the quark field operator and 
 $\lambda_{a}$ the octet of SU(3) matrices. To generate the relevant 
 interactions, $\psi$ is expanded in creation and annihilation 
 operators (for charm, up, down, and strange quarks), but transitions 
 from two mesons to three mesons and all transitions that violate the 
 Zweig rule are omitted. It is a good approximation to neglect all 
 effects of the Coulomb piece of the potential in (\ref{eq:CCCMH}).

 \newcolumntype{+}{D{:}{ \pm }{-4}}
 
 A full outline of the calculational procedure appears in
 Refs.~\cite{Eichten:1978tg,Eichten:1980ms}, but it is apt to cite a
 few elements here.  We evaluate Eq.~\ref{eq:CCCMH} between 
 nonrelativistic $(c\bar{c})$ states with wave functions determined 
 by the Cornell potential, and 1\slj{1}{1}{0} and 1\slj{3}{1}{1} 
 $c\bar{u}$, $c\bar{d}$, and $c\bar{s}$ ground states with Gaussian 
 wave functions. 
 States with orbital angular momentum $L > 0$ can decay in partial 
 waves $\ell = L \mp 1$.
 
 Following~\cite{Eichten:1978tg}, we define a coupling matrix within
 the $(c\bar{c})$ sector
 \begin{equation}
     \Omega_{nm}(W)= \sum_{ij} 
     \frac{\bra{n} \mathcal{H}_{I}\ket{D_{i}\bar{D}_{j}} 
     \bra{D_{i}\bar{D}_{j}} \mathcal{H}_{I} \ket{m}}
     {(W-E_{D_{i}}-E_{\bar{D}_{j}} + i\varepsilon)}\; ,
     \label{eq:omeag}
 \end{equation}
 where the summation  runs over momentum, spin, and flavor. 
 Above threshold (for $W > M_{i} + M_{j}$), $\Omega$ is complex. We 
 decompose $\Omega_{nm}$ into a dynamical part 
 (see~\cite{Eichten:1978tg}) that depends on the radial and orbital 
 quantum numbers of the charmonium states and on the masses of 
 $D_{i}$ and $D_{j}$ times the product recoupling matrix shown in
 Table~\ref{table:statcoups} that expresses the spin dependence for 
 each partial wave.

 \begin{table}
     \caption{Statistical recoupling coefficients $C$, defined by 
     Eq.~D19 of Ref.~{\protect\cite{Eichten:1978tg}}, that enter the calculation 
     of charmonium decays to pairs of charmed mesons. Paired entries 
     correspond to $\ell = L-1$ and $\ell = L+1$. \label{table:statcoups}}
 \begin{ruledtabular}
     \begin{tabular}{cccc}
	  State & $D\bar{D}$ & $D\bar{D}^{*}$ & $D^{*}\bar{D}^{*}$  \\
	 \hline
	\slj{1}{1}{0} & -- : $0$ & -- : $2$ & -- : $2$  \\[3pt]
	 \slj{3}{1}{1} & -- : $\cfrac{1}{3}$ & -- : 
	 $\cfrac{4}{3}$ & -- : $\cfrac{7}{3}$  \\[3pt]
	 \slj{3}{2}{0} & $1$ : $0$ & $0$ : $0$ & $\cfrac{1}{3}$ : 
	 $\cfrac{8}{3}$  \\[3pt]
	 \slj{3}{2}{1} & $0$ : $0$ & $\cfrac{4}{3}$ : 
	 $\cfrac{2}{3}$ & $0$ : $2$  \\[3pt]
	 \slj{1}{2}{1} & $0$ : $0$ & $\cfrac{2}{3}$ : $\cfrac{4}{3}$ & 
	 $\cfrac{2}{3}$ : $\cfrac{4}{3}$   \\[3pt]
	 \slj{3}{2}{2} & $0$ : $\cfrac{2}{5}$ & $0$ : 
	 $\cfrac{6}{5}$ & $\cfrac{4}{3}$ : $\cfrac{16}{15}$  \\[3pt]
	 \slj{3}{3}{1} & $\cfrac{2}{3}$ : $0$ & $\cfrac{2}{3}$ : 
	 $0$ & $\cfrac{4}{15}$ : $\cfrac{12}{5}$  \\[3pt]
	 \slj{3}{3}{2} & $0$ : $0$ & $\cfrac{6}{5}$ : 
	 $\cfrac{4}{5}$ & $\cfrac{2}{5}$ : $\cfrac{8}{5}$  \\[3pt]
	 \slj{1}{3}{2} & $0$ : $0$ & $\cfrac{4}{5}$ : $\cfrac{6}{5}$ & 
	 $\cfrac{4}{5}$ : $\cfrac{6}{5}$  \\[3pt]
	 \slj{3}{3}{3} & $0$ : $\cfrac{3}{7}$ & $0$ : 
	 $\cfrac{8}{7}$ & $\cfrac{8}{5}$ : $\cfrac{29}{35}$  \\[3pt]
	 \slj{3}{4}{2} & $\cfrac{3}{5}$ : $0$ & $\cfrac{4}{5}$ : 
	 $0$ & $\cfrac{11}{35}$ : $\cfrac{16}{7}$  \\[3pt]
	 \slj{3}{4}{3} & $0$ : $0$ & $\cfrac{8}{7}$ : 
	 $\cfrac{6}{7}$ & $\cfrac{4}{7}$ : $\cfrac{10}{7}$  \\[3pt]
	 \slj{1}{4}{3} & $0$ : $0$  & $\cfrac{6}{7}$ : $\cfrac{8}{7}$ & 
	 $\cfrac{6}{7}$ : $\cfrac{8}{7}$   \\[3pt]
	 \slj{3}{4}{4} & $0$ : $\cfrac{4}{9}$ & $0$ : 
	 $\cfrac{10}{9}$ & $\cfrac{12}{7}$ : $\cfrac{46}{63}$  \\[3pt]
	 \slj{3}{5}{3} & $\cfrac{4}{7}$ : $0$ & $\cfrac{6}{7}$ : 
	 $0$ & $\cfrac{22}{63}$ : $\cfrac{20}{9}$  \\[3pt]
	 \slj{3}{5}{4} & $0$ : $0$ & $\cfrac{10}{9}$ : 
	 $\cfrac{8}{9}$ & $\cfrac{2}{3}$ : $\cfrac{4}{3}$  \\[3pt]
	 \slj{1}{5}{4} & $0$ : $0$ & $\cfrac{8}{9}$ : $\cfrac{10}{9}$ & 
	 $\cfrac{8}{9}$ : $\cfrac{10}{9}$  \\[3pt]
	 \slj{3}{5}{5} & $0$ : $\cfrac{5}{11}$ & $0$ : 
	 $\cfrac{12}{11}$ & $\cfrac{16}{9}$ : $\cfrac{67}{99}$  \\
     \end{tabular}
 \end{ruledtabular}
 \end{table}

 In each channel $\sLj{2S+1}{L}{J}$, the physical states correspond to the
 eigenvalues of
 \begin{equation}
     \left(\mathcal{H}_{c\bar{c}} + \Omega(W)\right)\Psi = W\Psi\;.
     \label{eq:eigen}
 \end{equation}
 The real parts of the energy eigenvalues 
 are the charmonium masses. Imaginary parts determine the widths of 
 resonances above threshold. The eigenvalues also determine the 
 mixing among $(c\bar{c})$ states and the overall fraction in the 
 $(c\bar{c})$ sector.
 
 To fix the (Coulomb $+$ linear) charmonium potential, 
 \begin{equation}
     V(r) = -\kappa/r + r/a^{2},
     \label{eq:coulin}
 \end{equation}
 we adjust the strength of the linear term to reproduce the observed
 $\psi^{\prime}$-$\psi$ splitting, \textit{after including all the
 effects of coupling to virtual decay channels.} Neglecting the 
 influence of open charm gives $a=2.34\gev$, $\kappa=0.52$, and a 
 charmed-quark mass $m_{c} = 1.84\gev$. In the Cornell 
 coupled-channel model, the virtual decay channels reduce the 
 $\psi^{\prime}$-$\psi$ splitting by about $115\mev$, so the slope 
 parameter has to be reduced to $a = 1.97\gev$.

 \begin{table}
 \caption{Charmonium spectrum, including the influence of open-charm 
 channels. All masses are in MeV. The penultimate column holds an estimate 
 of the spin splitting due to tensor and spin-orbit forces in a 
 single-channel potential model. The last 
 column gives the spin splitting induced by communication with 
 open-charm states, for an initially unsplit multiplet.
 \label{table:delM}}
 \begin{ruledtabular}
 \begin{tabular}{ccccc}
 State & Mass & Centroid & $\begin{array}{c} \text{Splitting} 
 \\ \text{(Potential)} \end{array}$ &  $\begin{array}{c} \text{Splitting} 
 \\ \text{(Induced)} \end{array}$ \\
 \hline
 & & & & \\[-9pt]
 $\begin{array}{c}
     1\slj{1}{1}{0} \\
     1\slj{3}{1}{1} 
 \end{array}$ &
 $\begin{array}{c} 2\,979.9\footnotemark[1]  \\ 3\,096.9\footnotemark[1]  
 \end{array}$ & $3\,067.6\footnotemark[2]$ & $\begin{array}{c} -90.5 
 \\ +30.2 \end{array}$
    &  $\begin{array}{c} +2.8  \\ -0.9 \end{array} $\\ & & & & \\[-6pt]
 $\begin{array}{c} 
 1\slj{3}{2}{0} \\
 1\slj{3}{2}{1} \\
 1\slj{1}{2}{1} \\
 1\slj{3}{2}{2}  \end{array}$ &  $\begin{array}{c}
 3\,415.3\footnotemark[1]\\
 3\,510.5\footnotemark[1]\\
 3\,525.3\\
 3\,556.2\footnotemark[1] \end{array}$
 & $3\,525.3\footnotemark[3]$ &  $\begin{array}{c} 
 -114.9\footnotemark[5] \\ -11.6\footnotemark[5] 
 \\ +1.5\footnotemark[5] \\ -31.9\footnotemark[5] \end{array}$ & 
 $\begin{array}{c} +5.9 \\ -2.0 \\ +0.5 \\ -0.3
 \end{array}$ \\
 & & & & \\[-6pt]
 $\begin{array}{c}
     2\slj{1}{1}{0} \\
     2\slj{3}{1}{1} 
 \end{array}$ &
 $\begin{array}{c} 3\,637.7\footnotemark[1] \\ 3\,686.0\footnotemark[1]
 \end{array}$ & $3\,673.9\footnotemark[2]$ & $\begin{array}{c} -50.4 
 \\ +16.8 \end{array}$
    &  $\begin{array}{c} +15.7  \\ -5.2 \end{array} $\\
    & & & & \\[-6pt]
 $\begin{array}{c} 
 1\slj{3}{3}{1} \\
 1\slj{3}{3}{2} \\
 1\slj{1}{3}{2} \\
 1\slj{3}{3}{3}  \end{array}$ &  $\begin{array}{c}
 3\,769.9\footnotemark[1]\footnotemark[2]\\
 3\,830.6\\
 3\,838.0\\
 3\,868.3 \end{array}$
 & 
 (3\,815)\footnotemark[4] & 
 $\begin{array}{c} -40 \\ 0 \\ 
 0 \\ +20 \end{array} $ & 
 $\begin{array}{c} -39.9 \\ -2.7\\ +4.2 \\ +19.0
 \end{array}$ \\ & & & & \\[-6pt]
 $\begin{array}{c} 
 2\slj{3}{2}{0} \\
 2\slj{3}{2}{1} \\
 2\slj{1}{2}{1} \\
 2\slj{3}{2}{2}  \end{array}$ &  $\begin{array}{c}
 3\,931.9\\
 4\,007.5\\
 3\,968.0\\
 3\,966.5 \end{array}$
 & 3\,968\footnotemark[4] & 
 $\begin{array}{c} -90 \\ -8 \\ 
 0 \\ +25 \end{array} $ & 
 $\begin{array}{c} +10 \\ +28.4 \\ -11.9 \\ -33.1
 \end{array}$ \\
 \end{tabular}
 \end{ruledtabular}
 \footnotetext[1]{Observed mass, from \textit{Review of Particle 
 Physics,} Ref.~\cite{PDBook}.}
 \footnotetext[2]{Inputs to potential determination.}
 \footnotetext[3]{Observed 1\slj{3}{2}{J} centroid.}
 \footnotetext[4]{Computed.}
 \footnotetext[5]{Required to reproduce observed masses.}
 \end{table}
 
 The basic coupled-channel interaction (\ref{eq:CCCMH}) is
 spin-independent, but the hyperfine splittings of $D$ and $D^{*}$, 
 $D_{s}$ and $D_{s}^{*}$, induce spin-dependent forces that affect the
 charmonium states.  These spin-dependent forces give rise to S-D
 mixing that contributes to the $\psi(3770)$ electronic width, for
 example, and are a source of additional spin splitting, shown in the
 rightmost column of Table~\ref{table:delM}.  To compute the induced 
 splittings, we adjust the bare centroid of the spin-triplet states so 
 that the physical centroid, after inclusion of coupled-channel 
 effects, matches the value in the middle column of 
 Table~\ref{table:delM}. 
 As expected, the shifts
 induced in the low-lying 1S and 1P levels are small.  For the other
 known states in the 2S and 1D families, coupled-channel effects are
 noticeable and interesting.  

 In a simple potential picture, the $\eta_{c}^{\prime}(2\slj{1}{1}{0})$ 
 level lies below the $\psi^{\prime}(2\slj{3}{1}{1})$ by the hyperfine 
 splitting given by $M(\psi^{\prime}) - M(\eta_{c}^{\prime}) =
 32\pi\alpha_{s}|\Psi(0)|^{2}/9m_{c}^{2}$. Normalizing to the observed 
 1S hyperfine splitting, $M(\jpsi) - M(\eta_{c}) = 117\mev$, we  
 would find 
  \begin{equation}
     M(\psi^{\prime}) - M(\eta_{c}^{\prime}) = 67\mev\; ,
 \end{equation}
which is larger than the observed $48.3 \pm 4.4\mev$, as is typical 
for potential-model calculations.  The 2S induced
 shifts in Table~\ref{table:delM} draw $\psi^{\prime}$ and
 $\eta_{c}^{\prime}$ closer by $20.9\mev$, substantially improving the
 agreement between theory and experiment.  It is tempting to conclude 
 that the $\psi^{\prime}$-$\eta_{c}^{\prime}$ splitting reflects the 
 influence of virtual decay channels.
 
We lack a comprehensive theory of spin 
 splittings for $L>0$ states, and various potential-model schemes differ 
 appreciably in their predictions. (See Table I of 
 Ref.~\cite{Barnes:2003vb} for a variety of estimates.) For the 1P 
 states, the spin splittings shown under Splitting 
 (Potential) in Table~\ref{table:delM}
  are those required to 
 reproduce the observed masses; they are not predictions. For the 1D 
 and 2P levels, we have 
 adopted as representative the spin splittings shown.

 To reproduce the observed mass of the 1\slj{3}{3}{1} $\psi(3770)$, we
 shift the bare 1D centroid upward by $67.5\mev$.  The other 1D masses 
 are thus pegged to the observed $\psi(3770)$. In our model calculation, the
 coupling to open-charm
 channels increases the 1\slj{3}{3}{2}-1\slj{3}{3}{1} splitting by
 about $20\mev$, but does not fully account for the observed $102\mev$
 separation between $X(3872)$ and $\psi(3770)$.  It is noteworthy that
 the position of the $3^{--}$ 1\slj{3}{3}{3} level turns out to be very
 close to $3872\mev$.  For the 2P levels, we have no experimental 
 anchor, so we adjust the bare centroid so that the 2\slj{1}{2}{1} 
 level lies at the centroid of the potential-model calculation. 
 It is likely that we have more to learn about
 the influence of open-charm channels.
 
 The 2\slj{1}{2}{1} level has been suggested~\cite{Pakvasa:2003ea} as 
 an alternative assignment for $X(3872)$ because it has an allowed 
 $\pi\pi$ transition to $\jpsi$ and a hindered M1 radiative transition to 
 the 1P levels. The coupled-channel calculation places this state 
 nearly $100\mev$ above $D\bar{D}^{*}$ threshold. As we shall see in 
 quantitative detail presently, its allowed $s$-wave decay to 
 $D^{0}\bar{D}^{*0}$ leads to an unacceptably large width, unless 
 $X(3872)$ lies below $D^{0}\bar{D}^{*0}$ threshold.

The wave functions that correspond to physical states are linear
combinations of potential-model $c\bar{c}$ eigenstates plus admixtures
of charmed-meson pairs.  We record the charmonium content of states of
interest in Table~\ref{table:wavefcns}.  The open-charm pieces have the
spatial structure of bound states of charmed mesons, but they are not
molecular charm states in the usual sense: they are virtual
contributions for states below threshold, and---unlike ``deusons,'' for
example~\cite{Tornqvist:2004qv}---they are not bound by one-pion
exchange.
 
\begin{table*}
     \caption{Charmonium content of states near flavor threshold. 
     The wave function $\Psi$ takes account of mixing induced through
     open charm-anticharm channels. Unmixed potential-model 
     eigenstates are denoted by $\ket{n\sLj{2s+1}{L}{J}}$. The 
     coefficient of the dominant eigenstate is chosen real and 
     positive. The 1S, 1P, 2S, and 1\slj{3}{3}{1} states are 
     evaluated at their physical masses. The remaining 1D and 2P 
     states are considered at the potential-model centroids \textbf{or 
     at the masses in Table~\ref{table:delM}?}. We 
     also show the  1\slj{3}{3}{2}, 1\slj{3}{3}{3}, and 
     2\slj{1}{2}{1} states at the mass of $X(3872)$. $\zcc$ 
     represents the $(c\bar{c})$ probability fraction of each state. 
     \label{table:wavefcns}}
 \begin{ruledtabular}
  \begin{tabular}{l}
$\Psi(1\slj{1}{1}{0}) = 0.986\ket{1\slj{1}{1}{0}} - 0.042\ket{2\slj{1}{1}{0}}
  -0.008\ket{3\slj{1}{1}{0}} -0.002\ket{4\slj{1}{1}{0}}
  -0.001\ket{5\slj{1}{1}{0}}; \quad 
  \zcc = 0.974$ \\
$\Psi(1\slj{3}{1}{1}) = 0.983\ket{1\slj{3}{1}{1}} -0.050\ket{2\slj{3}{1}{1}}
-0.009\ket{3\slj{3}{1}{1}} - 0.003\ket{4\slj{3}{1}{1}} + 
-0.001\ket{5\slj{3}{1}{1}}; \quad \zcc=0.968$\\[3pt]
$\Psi(1\slj{3}{2}{0}) = 0.919\ket{1\slj{3}{2}{0}} -0.067\ket{2\slj{3}{2}{0}}
-0.014\ket{3\slj{3}{2}{0}} - 0.005\ket{4\slj{3}{2}{0}}
-0.002\ket{5\slj{3}{2}{0}}; \quad \zcc = 
0.850$\\
$\Psi(1\slj{3}{2}{1}) = 0.914\ket{1\slj{3}{2}{1}} -0.075\ket{2\slj{3}{2}{1}}
-0.015\ket{3\slj{3}{2}{1}} -0.005\ket{4\slj{3}{2}{1}}
-0.002\ket{5\slj{3}{2}{1}}; \quad \zcc = 
0.841$ \\
$\Psi(1\slj{1}{2}{1}) = 0.918\ket{1\slj{1}{2}{1}} - 0.077\ket{2\slj{1}{2}{1}}
-0.015\ket{3\slj{1}{2}{1}} -0.005\ket{4\slj{1}{2}{1}}
-0.002\ket{5\slj{1}{2}{1}}; \quad 
\zcc=0.845$ \\
$\Psi(1\slj{3}{2}{2}) = 0.920\ket{1\slj{3}{2}{2}} -0.080\ket{2\slj{3}{2}{2}}
-0.015\ket{3\slj{3}{2}{2}} -0.005\ket{4\slj{3}{2}{2}} 
-0.002\ket{5\slj{3}{2}{2}}
-0.002\ket{1\slj{3}{4}{2}}; \quad \zcc = 0.854$ \\[6pt]
$\Psi(2\slj{1}{1}{0}) = 0.087\ket{1\slj{1}{1}{0}} + 0.883\ket{2\slj{1}{1}{0}}
-0.060\ket{3\slj{1}{1}{0}} -0.016\ket{4\slj{1}{1}{0}} 
-0.007\ket{5\slj{1}{1}{0}} - 0.003\ket{6\slj{1}{1}{0}}; \quad 
\zcc=0.791$ \\
$\Psi(2\slj{3}{1}{1}) = 0.103\ket{1\slj{3}{1}{1}} + 0.838\ket{2\slj{3}{1}{1}}
-0.085\ket{3\slj{3}{1}{1}} - 0.017\ket{4\slj{3}{1}{1}} 
-0.007\ket{5\slj{3}{1}{1}} - 0.002\ket{6\slj{3}{3}{1}}$ \\
\quad
$+0.040\ket{1\slj{3}{3}{1}} -0.008\ket{2\slj{3}{3}{1}}; \quad 
\zcc=0.723$\\[6pt]
$\Psi(1\slj{3}{3}{1}) = 0.694\ket{1\slj{3}{3}{1}} +0.097\,e^{0.935i\pi}\ket{2\slj{3}{3}{1}} 
+ 0.008\,e^{-0.668i\pi}\ket{3\slj{3}{3}{1}} + 0.006\,e^{0.904i\pi}\ket{4\slj{3}{3}{1}}
$\\ 
\quad
$+0.013\,e^{0.742i\pi}\ket{1\slj{3}{1}{1}} +0.168\,e^{0.805i\pi}\ket{2\slj{3}{1}{1}}+ 0.014\,e^{0.866i\pi}\ket{3\slj{3}{1}{1}} 
+ 0.012\,e^{-0.229i\pi}\ket{4\slj{3}{1}{1}}$ \\ \quad\quad
$+0.001\,e^{0.278i\pi}\ket{5\slj{3}{1}{1}} + 
0.001\,e^{-0.267i\pi}\ket{6\slj{3}{1}{1}}; \quad\zcc=0.520$\\
$\Psi(1\slj{3}{3}{2}) = 0.754\ket{1\slj{3}{3}{2}} -0.084\ket{2\slj{3}{3}{2}}
-0.011\ket{3\slj{3}{3}{2}} -0.006\ket{4\slj{3}{3}{2}}; \quad 
\zcc=0.576$ \\
$\Psi(1\slj{1}{3}{2}) = 0.770\ket{1\slj{1}{3}{2}} -0.083\ket{2\slj{1}{3}{2}}
-0.012\ket{3\slj{1}{3}{2}} -0.006\ket{4\slj{1}{3}{2}}; \quad 
\zcc=0.600$ \\
$\Psi(1\slj{3}{3}{3}) = 0.812\ket{1\slj{3}{3}{3}} 
+0.086\,e^{0.990i\pi}\ket{2\slj{3}{3}{3}} +0.013\,e^{-0.969i\pi}\ket{3\slj{3}{3}{3}}
+0.007\,e^{0.980i\pi}\ket{4\slj{3}{3}{3}}$\\
\quad $+0.016\,e^{0.848i\pi}\ket{1\slj{3}{5}{3}}
+0.003\,e^{-0.291i\pi}\ket{2\slj{3}{5}{3}}; \quad \zcc=0.667$ \\[6pt]
$\Psi(2\slj{3}{2}{0}) = 0.040\,e^{-0.454i\pi}\ket{1\slj{3}{2}{0}} + 
0.532\ket{2\slj{3}{2}{0}} + 0.024\,e^{-0.889i\pi}\ket{3\slj{3}{2}{0}}
+ 0.010\,e^{0.867i\pi}\ket{4\slj{3}{2}{0}}$ \\ \quad
$+0.006\,e^{-0.976i\pi}\ket{5\slj{3}{2}{0}}; \quad \zcc=0.286$ \\
$\Psi(2\slj{3}{2}{1}) = 0.218\,e^{-0.456i\pi}\ket{1\slj{3}{2}{1}} + 
0.821\ket{2\slj{3}{2}{1}} + 0.058\,e^{0.516i\pi}\ket{3\slj{3}{2}{1}}
+0.032\,e^{0.976i\pi}\ket{4\slj{3}{2}{1}}$\\ \quad$ + 
0.008\,e^{0.986i\pi}\ket{5\slj{3}{2}{1}}; \quad \zcc=0.726$\\
$\Psi(2\slj{1}{2}{1}) = 0.216\,e^{-0.226i\pi}\ket{1\slj{1}{2}{1}} +
0.852\ket{2\slj{1}{2}{1}} +0.079\,e^{0.780i\pi}\ket{3\slj{1}{2}{1}}
+0.023\,e^{-0.890i\pi}\ket{4\slj{1}{2}{1}}$ \\
\quad $0.007\,e^{0.985i\pi}\ket{5\slj{1}{2}{1}}; \quad \zcc = 0.883$ \\
$\Psi(2\slj{3}{2}{2}) = 0.234\,e^{-0.046i\pi}\ket{1\slj{3}{2}{2}}
+0.754\ket{2\slj{3}{2}{2}} + 0.097\,e^{0.876i\pi}\ket{3\slj{3}{2}{2}}
+0.016\,e^{-0.743i\pi}\ket{4\slj{3}{2}{2}}$\\
\quad $0.007\,e^{0.898i\pi}\ket{5\slj{3}{2}{2}}
+0.370\,e^{0.775i\pi}\ket{1\slj{3}{4}{2}} + 0.035\,e^{-0.317i\pi}\ket{2\slj{3}{4}{2}}
+ 0.002\,e^{0.097i\pi}\ket{3\slj{3}{4}{2}}; \quad \zcc=0.771$\\[9pt]
$M=3872\mev:\;\Psi(1\slj{3}{3}{2}) = 0.596\ket{1\slj{3}{3}{2}} -0.108\ket{2\slj{3}{3}{2}}
-0.004\ket{3\slj{3}{3}{2}} -0.006\ket{4\slj{3}{3}{2}}; \quad 
\zcc=0.367$ \\
$M=3872\mev:\;\Psi(1\slj{3}{3}{3}) = 0.813\ket{1\slj{3}{3}{3}} 
+0.089\,e^{0.989i\pi}\ket{2\slj{3}{3}{3}} +0.013\,e^{-0.965i\pi}\ket{3\slj{3}{3}{3}}
+0.007\,e^{0.978i\pi}\ket{4\slj{3}{3}{3}}$\\
\phantom{$M=3872\mev:\;\Psi(1\slj{3}{3}{1})$}
$+0.017\,e^{0.837i\pi}\ket{1\slj{3}{5}{3}}
+0.003\,e^{-0.305i\pi}\ket{2\slj{3}{5}{3}}; \quad \zcc=0.669$ \\[3pt]
$M=3872\mev:\;\Psi(2\slj{1}{2}{1}) = 0.134\,e^{-0.004i\pi}\ket{1\slj{1}{2}{1}}
+0.374\ket{2\slj{1}{2}{1}} + 0.035\,e^{0.993i\pi}\ket{3\slj{1}{2}{1}}
+0.003\,e^{-0.981i\pi}\ket{4\slj{1}{2}{1}} $ \\ \quad
$+0.004\,e^{0.996i\pi}\ket{5\slj{1}{2}{1}}; \quad \zcc=0.159$
  \end{tabular}
  \end{ruledtabular}
  \end{table*}

\textit{Expectations for radiative transitions.} As Table~\ref{table:wavefcns} 
shows, the physical charmonium states are not pure potential-model 
eigenstates. To compute the E1 radiative transition rates, we must 
take into account both the standard $(c\bar{c}) \to (c\bar{c})\gamma$ 
transitions and the transitions between (virtual) decay channels in 
the initial and final states. Details of the calculational procedure 
are given in \S IV.B of Ref.~\cite{Eichten:1980ms}. 

Our expectations for E1 transition rates among spin-triplet levels are
shown in Table~\ref{table:radtranstw}.  There we show both the rates
calculated between single-channel potential-model eigenstates (in
italics) and the rates that result from the Cornell coupled-channel
model, to indicate the influence of the open-charm channels.
The model reproduces the trends of transitions to and from the $\chi_{c}$ 
states in broad outline. Not 
surprisingly, the single-channel values roughly track those calculated by 
Barnes \& Godfrey in their potential~\cite{Barnes:2003vb}. For
these low-lying states, the mixing through open-charm channels results
in a mild reduction of the rates.

We show the 1\slj{3}{3}{1} transition rates at the mass of $\psi(3770)$
and at the predicted 1\slj{3}{3}{1} centroid, $3815\mev$.  For the
$\psi(3770)$, with its total width of about $24\mev$, the
$1\slj{3}{3}{1}(3770) \to \chi_{c0}\,\gamma(338)$ transition might
someday be observable with a branching fraction of 1\%.

For the 1\slj{3}{3}{2} and 1\slj{3}{3}{3} levels, we have computed the
radiative decay rates at the predicted 1\slj{3}{3}{1} centroid,
$3815\mev$, at the mass calculated for the states ($3831\mev$ and
$3868\mev$, respectively), and at the mass of $X(3872)$.  We will
compare the partial widths for the $\chi_{c1}\,\gamma(344)$ and
$\chi_{c2}\,\gamma(303)$ with the expected $\pi^{+}\pi^{-}\jpsi$ and
open-charm decay rates presently.

We have evaluated the radiative decay rates for the 2\slj{3}{2}{J} 
levels at the calculated centroid and at the predicted mass, where 
that is displaced appreciably from the centroid. We shall see below 
that all of these rates are small compared to the expected open-charm 
decay rates. 

\begin{table}
\caption{Calculated and observed rates for E1 radiative transitions 
among charmonium levels. \textit{Values in italics} result if the influence of 
open-charm channels is not included.\label{table:radtranstw}}
\begin{ruledtabular}
\begin{tabular}{lc+} 
    Transition    & \multicolumn{2}{c}{Partial width (keV)} \\
    ($\gamma$ energy in MeV)  & Computed & 
\multicolumn{1}{c}{Measured}\\
\hline
 & & \\[-6pt]
$\chi_{c0} \to\jpsi\,\gamma(303)$ & 
 $\mathit{113}\to 107$ & 119 : 19\footnotemark[1]  \\
$\chi_{c1} \to\jpsi\,\gamma(390)$ & 
 $\mathit{228} \to 216$  & 291 : 48\footnotemark[1] \\
$\chi_{c2} \to\jpsi\,\gamma(429)$ & 
 $\mathit{300} \to 287$  & 426 : 51\footnotemark[1] \\[6pt]
$\psi^{\prime} \to
\chi_{c2}\,\gamma(129)$ &  $\mathit{23} \to 23$ & 27 : 
4\footnotemark[2] \\
$\psi^{\prime} \to
\chi_{c1}\,\gamma(172)$ &  $\mathit{33} \to 32$ & 27 : 
3\footnotemark[2]   \\
$\psi^{\prime} \to
\chi_{c0}\,\gamma(261)$ &  $\mathit{36} \to 38$ & 27 : 
3\footnotemark[2]  \\[6pt]
$1\slj{3}{3}{1}(3770) \to
\chi_{c2}\,\gamma(208)$ &  $\text{\textit{3.2}} \to 3.9$ & \\
$1\slj{3}{3}{1}(3770)  \to
\chi_{c1}\,\gamma(251)$ &  $\mathit{183} \to 59$ & \\
$1\slj{3}{3}{1}(3770)  \to
\chi_{c0}\,\gamma(338)$ &  $\mathit{254} \to 225$ & \\[3pt]
$1\slj{3}{3}{1}(3815) \to
\chi_{c2}\,\gamma(250)$ &  $\text{\textit{5.5}} \to 6.8$ & \\
$1\slj{3}{3}{1}(3815)  \to
\chi_{c1}\,\gamma(293)$ &  $\mathit{128} \to 120$ & \\
$1\slj{3}{3}{1}(3815)  \to
\chi_{c0}\,\gamma(379)$ &  $\mathit{344} \to 371$ & \\[6pt]
$1\slj{3}{3}{2}(3815)\to\chi_{c2}\,\gamma(251)$ & 
 $\mathit{50} \to 40$ & \\
 $1\slj{3}{3}{2}(3815)\to\chi_{c1}\,\gamma(293)$ & 
  $\mathit{230} \to 191$ & \\[3pt]
  $1\slj{3}{3}{2}(3831)\to\chi_{c2}\,\gamma(266)$ & 
   $\mathit{59} \to 45$ & \\
   $1\slj{3}{3}{2}(3831)\to\chi_{c1}\,\gamma(308)$ & 
    $\mathit{264} \to 212$ & \\[3pt]
$1\slj{3}{3}{2}(3872)\to\chi_{c2}\,\gamma(303)$ & 
 $\mathit{85} \to 45$ & \\
 $1\slj{3}{3}{2}(3872)\to\chi_{c1}\,\gamma(344)$ & 
  $\mathit{362} \to 207$ & \\[6pt]
  $1\slj{3}{3}{3}(3815)\to\chi_{c2}\,\gamma(251)$ & 
   $\mathit{199} \to 179$ & \\[3pt]
   $1\slj{3}{3}{3}(3868)\to\chi_{c2}\,\gamma(303)$ & 
    $\mathit{329} \to 286$ & \\[3pt]
    $1\slj{3}{3}{3}(3872)\to\chi_{c2}\,\gamma(304)$ & 
     $\mathit{341} \to 299$ & \\[6pt]
$2\slj{3}{2}{0}(3933) \to \jpsi\,\gamma(747)$ & $\mathit{95} \to 19$ &
\\
$2\slj{3}{2}{0}(3933) \to \psi^{\prime}\,\gamma(239)$ & 
$\text{\textit{127}}
\to 38$ & \\
$2\slj{3}{2}{0}(3933) \to \psi(3770)\,\gamma(160)$ & $\mathit{59} \to 
     11$ & \\[3pt]
$2\slj{3}{2}{0}(3968) \to \jpsi\,\gamma(775)$ & $\mathit{110} \to 77$ & \\
$2\slj{3}{2}{0}(3968) \to \psi^{\prime}\,\gamma(272)$ & 
$\text{\textit{180}}
 \to 
155$ & \\
$2\slj{3}{2}{0}(3968) \to \psi(3770)\,\gamma(193)$ & $\mathit{101} \to 
43$ & \\[6pt]
$2\slj{3}{2}{1}(3968) \to \jpsi\,\gamma(775)$ & $\mathit{110} \to 68$ & \\
$2\slj{3}{2}{1}(3968) \to \psi^{\prime}\,\gamma(272)$ & 
$\text{\textit{180}} \to 
102$ & \\
$2\slj{3}{2}{1}(3968) \to \psi(3770)\,\gamma(193)$ & $\mathit{25} \to 
5$ & \\
$2\slj{3}{2}{1}(3968) \to 1\slj{3}{3}{2}(3815)\,\gamma(150)$ & 
$\mathit{37} \to 1.8$ &
\\
$2\slj{3}{2}{1}(3968) \to 1\slj{3}{3}{2}(3831)\,\gamma(135)$ & 
$\text{\textit{25}}
\to 0.25$ & \\
$2\slj{3}{2}{1}(3968) \to 1\slj{3}{3}{2}(3872)\,\gamma(95)$ & 
$\mathit{10} \to 
     0.23$ & \\[3pt]
     $2\slj{3}{2}{1}(4012) \to \jpsi\,\gamma(811)$ & $\mathit{132} \to 94$ &
     \\
     $2\slj{3}{2}{1}(4012) \to \psi^{\prime}\,\gamma(313)$ & 
     $\text{\textit{260}}
     \to 151$ & \\
     $2\slj{3}{2}{1}(4012) \to \psi(3770)\,\gamma(235)$ & $\mathit{43} \to 
	  11$ & \\[6pt]

     $2\slj{3}{2}{2}(3968) \to \jpsi\,\gamma(775)$ & $\mathit{110} \to 
     19$ & \\
     $2\slj{3}{2}{2}(3968) \to \psi^{\prime}\,\gamma(272)$ & 
     $\text{\textit{180}} \to 
     314$ & \\
     $2\slj{3}{2}{2}(3968) \to \psi(3770)\,\gamma(193)$ & $\text{\textit{1.0}} \to 
     1.4$ & \\
$2\slj{3}{2}{2}(3968) \to 1\slj{3}{3}{2}(3815)\,\gamma(150)$ &
$\text{\textit{7.4}} \to 18$ & \\
$2\slj{3}{2}{2}(3968) \to 1\slj{3}{3}{2}(3835)\,\gamma(131)$ &
$\mathit{5} \to 12$ & \\
$2\slj{3}{2}{2}(3968) \to 1\slj{3}{3}{2}(3872)\,\gamma(95)$ &
$\text{\textit{1.9}} \to 3.4$ & \\
$2\slj{3}{2}{2}(3968) \to 1\slj{3}{3}{3}(3815)\,\gamma(150)$ &
$\mathit{41} \to 82$ & \\
$2\slj{3}{2}{2}(3968) \to 1\slj{3}{3}{3}(3868)\,\gamma(99)$ &
$\mathit{12} \to 26$ & \\
$2\slj{3}{2}{2}(3968) \to 1\slj{3}{3}{3}(3872)\,\gamma(95)$ &
$\mathit{11} \to 23$ & \\[3pt]
\end{tabular}
\end{ruledtabular}
\footnotetext[1]{Derived from the 2003 ``unchecked fit'' of 
Ref.~\cite{PDBook}.}
\footnotetext[2]{Branching fractions from CLEO via 
Skwarnicki~\cite{Skwarnicki:2003wn}.}
\end{table}

\textit{Expectations for hadronic transitions.} The Beijing 
Spectrometer (BES)
observation~\cite{Bai:2003hv} of  a branching fraction $\mathcal{B}(\psi(3770) \to \pi^{+}\pi^{-}\jpsi)
= (0.59 \pm 0.26 \pm 0.16)\% $ would imply a hadronic cascade rate 
$\Gamma(1\slj{3}{3}{1} \to \pi\pi\jpsi) \approx 210 \pm 130\kev$, 
considerably larger than the $45\kev$, inferred~\cite{Eichten:1994gt} from older 
data, that we took as normalization in Ref.~\cite{Eichten:2002qv}. By 
the Wigner-Eckart theorem for E1-E1 transitions, all the 
$1\slj{3}{3}{J} \to \pi\pi\jpsi$ rates should be equal (for 
degenerate \slj{3}{3}{J} states), so the higher BES normalization would increase 
$\Gamma(1\slj{3}{3}{2} \to \pi\pi\jpsi)$ to hundreds of 
keV, as remarked by Barnes \& Godfrey~\cite{Barnes:2003vb}.  Combined 
with our estimate that $\Gamma(1\slj{3}{3}{2}(3872) \to \gamma 
\chi_{c1}) \approx 207\kev$, the larger $\pi\pi\jpsi$ rate relaxes 
somewhat---but does not eliminate---the tension between the \slj{3}{3}{2} 
assignment for $X$ and the Belle bound in Eq.~(\ref{eqn:radbound}).

The BES rate, which is based on a 
handful of events, is challenged by a CLEO-$c$ limit~\cite{Skwarnicki:2003wn}, 
$\mathcal{B}(\psi(3770) \to \pi^{+}\pi^{-}\jpsi) < 0.26\%$ at 90\% C.L. 
Both experiments are accumulating larger data samples that should 
improve our knowledge of this important normalization.

\section{Decays into open charm \label{sec:OCDK}} 
\begin{figure*}
    \BoxedEPSF{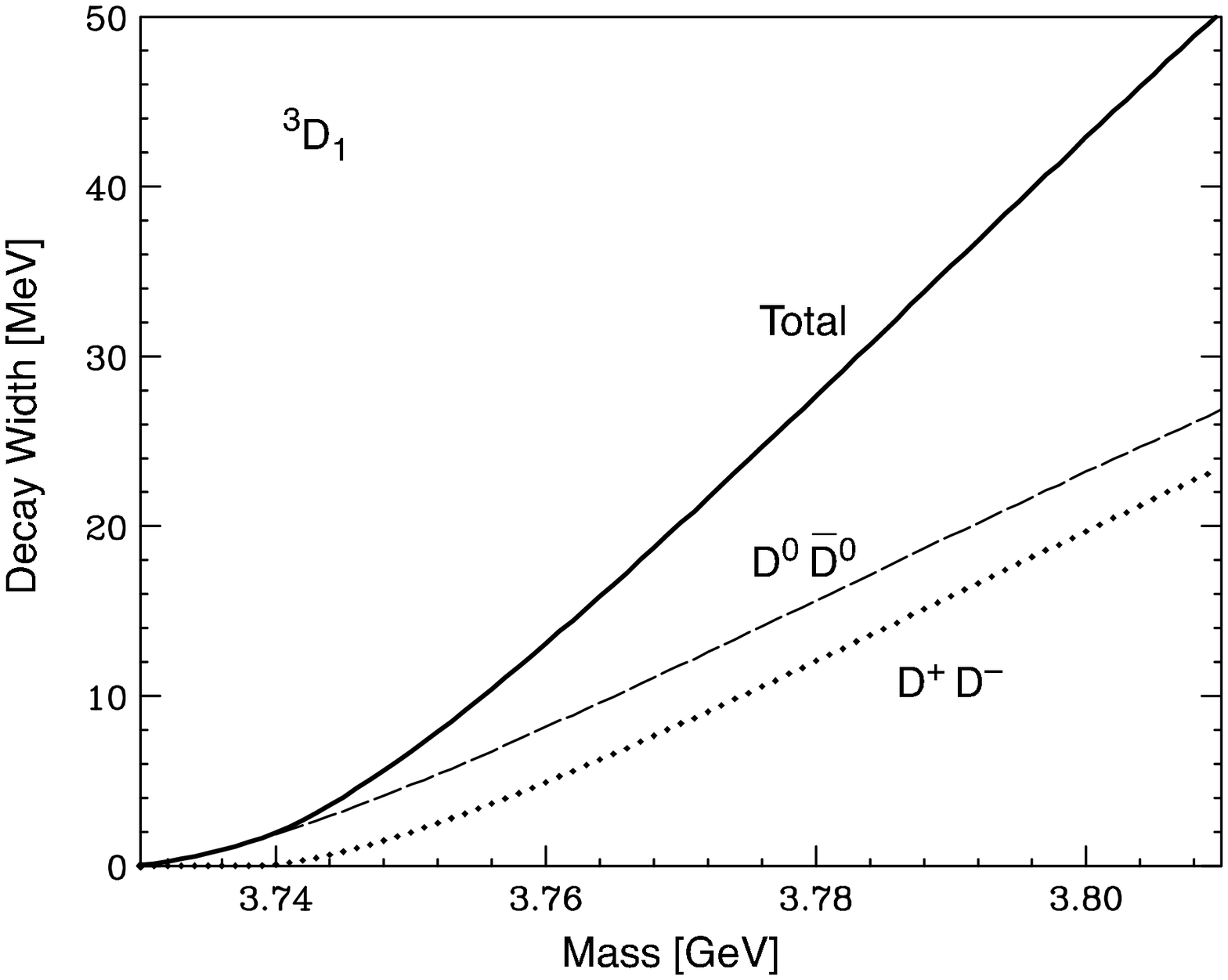 scaled 450}\quad\quad
    \BoxedEPSF{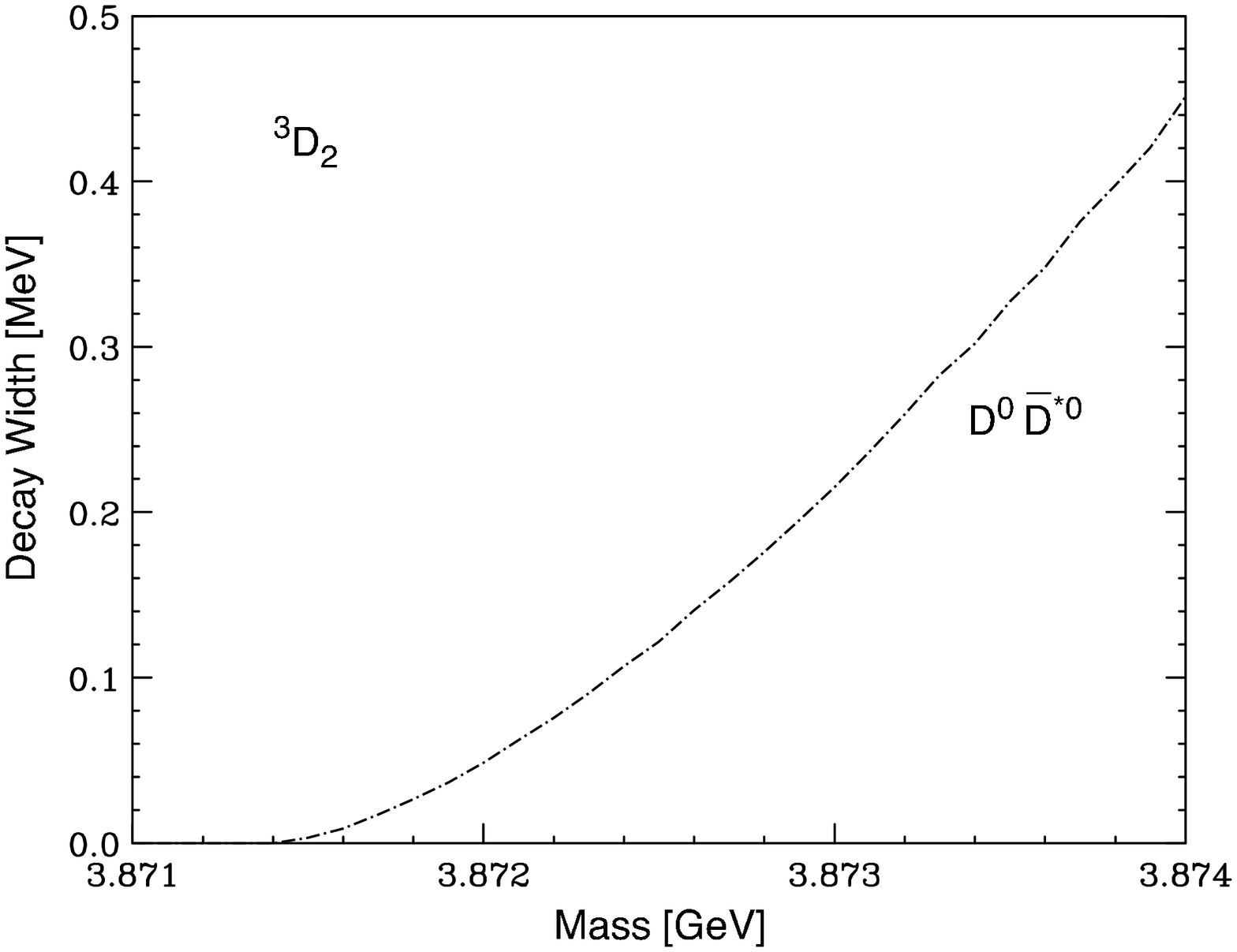 scaled 450}\\[3pt]
    \BoxedEPSF{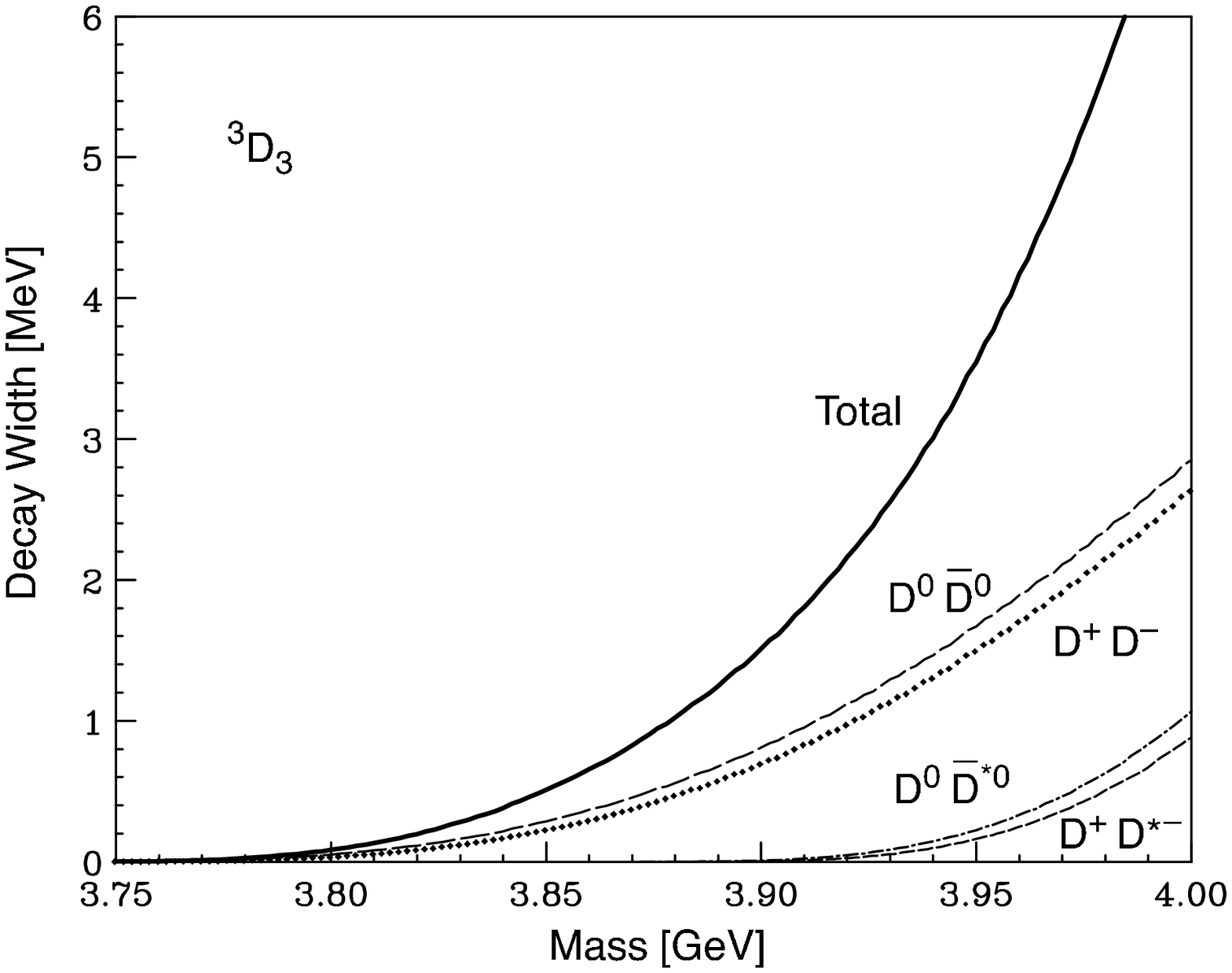 scaled 450}\quad\quad
    \BoxedEPSF{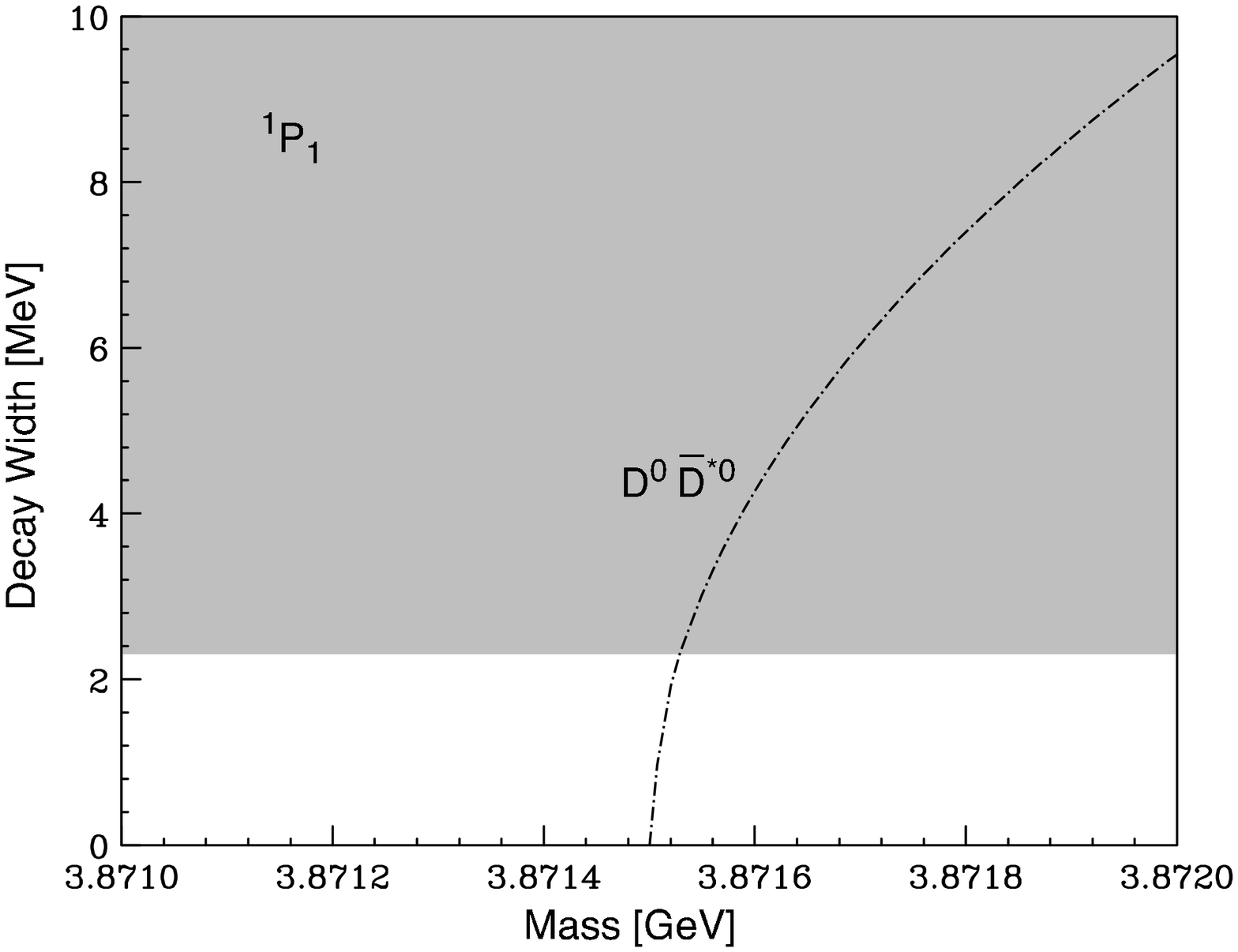 scaled 450}\\[3pt]
    \BoxedEPSF{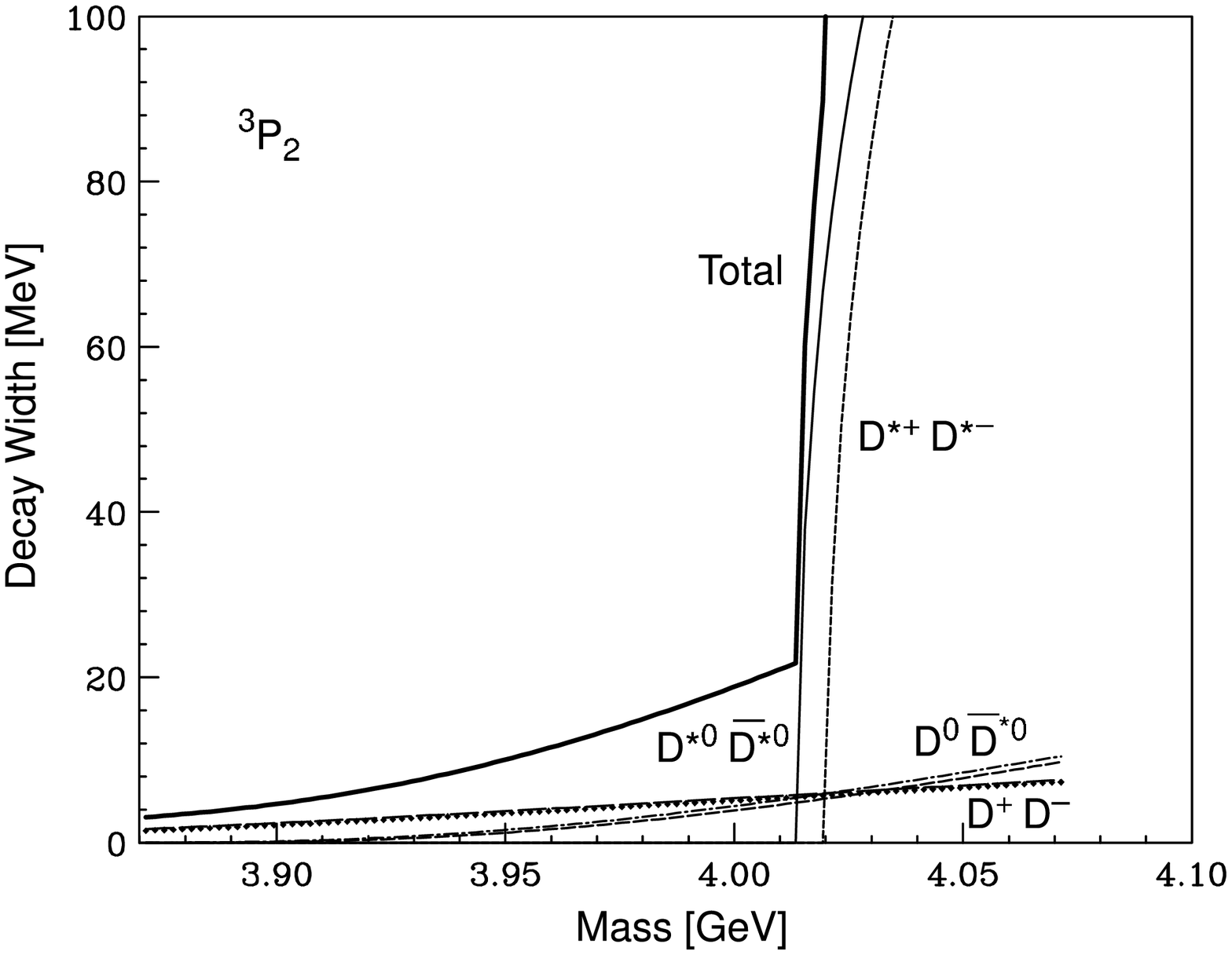 scaled 450}\quad\quad
   \BoxedEPSF{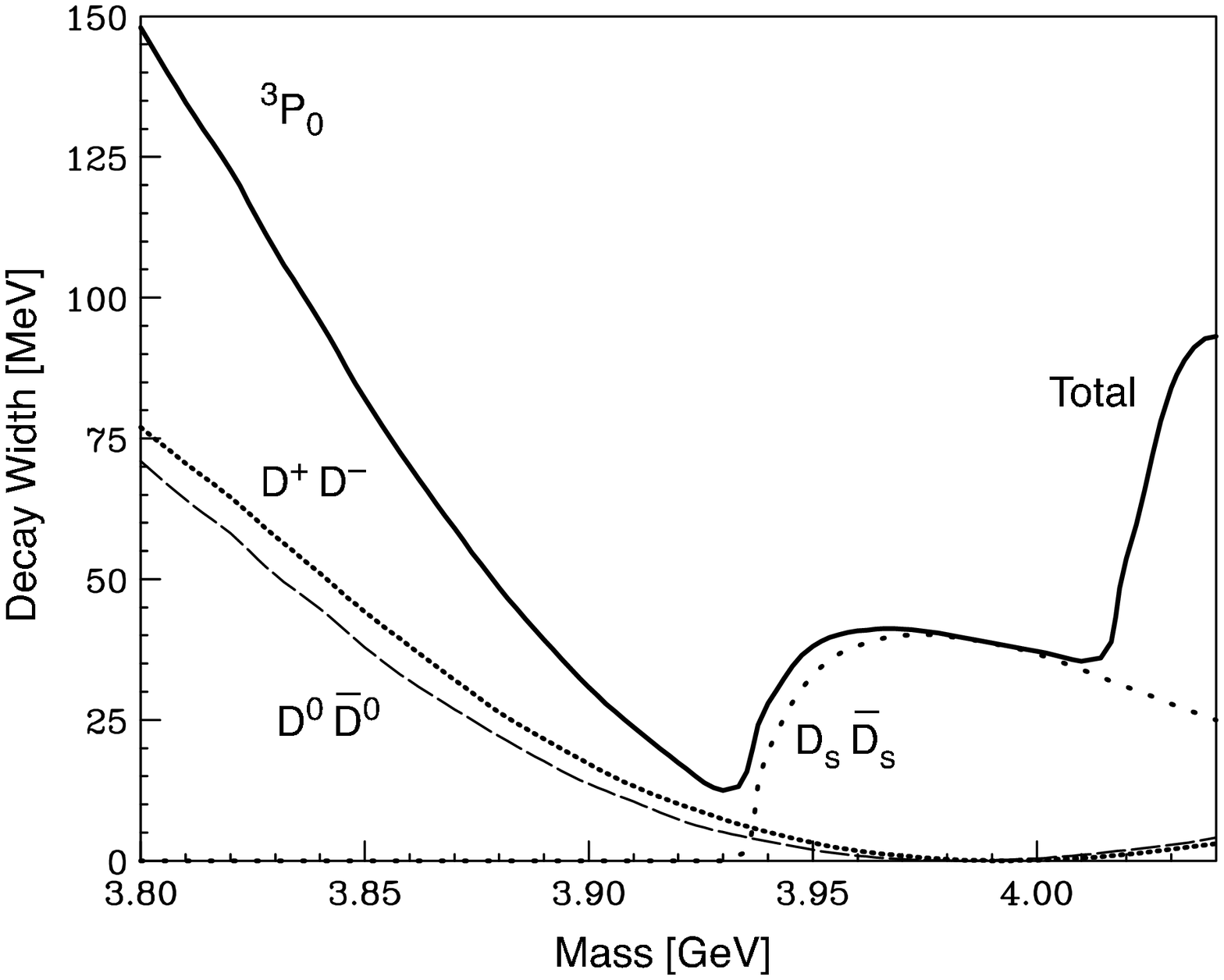 scaled 450}
    
    \caption{Partial and total widths near threshold for decay of 
    charmonium states into open charm, computed in the Cornell 
    coupled-channel model. Long dashes: $D^{0}\bar{D}^{0}$, dots: 
    $D^{+}D^{-}$, dot-dashes: $D^{0}\bar{D}^{*0}$, dashes: $D^{+}D^{*-}$, 
    thin line: $D^{*0}\bar{D}^{*0}$, short dashes: $D^{*+}D^{*-}$, 
    widely spaced dots: $D_{s}\bar{D}_{s}$,
    thick line: sum of open-charm channels. Belle's 90\% C.L.\ upper 
    limit~\cite{Choi:2003ue}, $\Gamma(X(3872)) < 2.3\mev$, is 
    indicated on the \slj{1}{2}{1} window. For $D\bar{D}^{*}$ modes, 
    the sum of $D\bar{D}^{*}$ and $\bar{D}D^{*}$ is always implied. \label{figure:OCdecays}}
\end{figure*}

The calculated partial widths for 
decays of charmonium states into open charm appear in 
Table~\ref{table:OCdecays}.  Experience~\cite{Eichten:1978tg} teaches 
that once the position of a resonance is given, the coupled-channel 
formalism yields reasonable predictions for the other resonance 
properties. The 1\slj{3}{3}{1} state $\psi^{\prime\prime}(3770)$, 
which lies some $40\mev$ above charm threshold,  
offers an important benchmark: we compute 
$\Gamma( \psi^{\prime\prime}(3770)\to 
D\bar{D}) = 20.1\mev$, to be compared with the Particle Data Group's 
fitted value of $23.6 \pm 2.7\mev$~\cite{PDBook}. The variation of the 
1\slj{3}{3}{1} width with mass is shown in the top left panel of 
Figure~\ref{figure:OCdecays}.  

Barnes \& Godfrey~\cite{Barnes:2003vb} have estimated the decays of
several of the charmonium states into open charm, using the
\slj{3}{2}{0} model of $q\bar{q}$ production first applied above charm 
threshold by the Orsay group~\cite{LeYaouanc:1977ux}.  They did not carry
out a coupled-channel analysis, and so did not determine the
composition of the physical states, but their estimates of open-charm 
decay rates can be read against ours as a rough assessment of model 
dependence.

The long-standing expectation that the 1\slj{3}{3}{2} and 1\slj{1}{3}{2} levels 
would be narrow followed from the presumption that these unnatural 
parity states should lie between the $D\bar{D}$ and $D\bar{D}^{*}$ 
thresholds, and could not decay into open charm. At $3872\mev$, both 
states can decay into $D^{0}\bar{D}^{*0}$, but the partial widths 
(Table~\ref{table:OCdecays}) are 
quite small. 
We show the variation of the 1\slj{3}{3}{2} partial width with mass in
the top right panel of Figure~\ref{figure:OCdecays}; over the region 
of interest, it does not threaten the Belle bound, 
$\Gamma(X(3872))<2.3\mev$. The range of values is quite similar to 
the range estimated for $\Gamma(1\slj{3}{3}{2} \to \pi\pi\jpsi)$, so 
we expect roughly comparable branching fractions for decays into 
$D^{0}\bar{D}^{*0}$ and $\pi^{+}\pi^{-}\jpsi$. If $X(3872)$ does turn 
out to be the 1\slj{3}{3}{2} level, we expect $M(1\slj{1}{3}{2}) = 
3880\mev$ and $\Gamma(1\slj{1}{3}{2} \to D^{0}\bar{D}^{*0}) \approx 
1.7\mev$.

The natural-parity 1\slj{3}{3}{3} state can decay into 
$D\bar{D}$, but its $f$-wave decay is suppressed by the centrifugal 
barrier factor, so the partial width is less than $1\mev$ at a mass 
of $3872\mev$. Although estimates of the hadronic cascade transitions 
are uncertain, the numbers in hand lead us to expect 
$\Gamma(1\slj{3}{3}{3} \to \pi^{+}\pi^{-}\jpsi) \alt \cfrac{1}{4} 
\Gamma(1\slj{3}{3}{3} \to D\bar{D})$, whereas $\Gamma(1\slj{3}{3}{3} 
\to \gamma\chi_{c2}) \approx \cfrac{1}{3}\Gamma(1\slj{3}{3}{3} \to 
D\bar{D})$, if $X(3872)$ is identified as 1\slj{3}{3}{3}.
The variation of $\Gamma(1\slj{3}{3}{3} \to 
D\bar{D})$ with mass is shown in the 
middle left panel of Figure~\ref{figure:OCdecays}. Note that if 
1\slj{3}{3}{3} is not to be identified with $X(3872)$, \textit{it may still 
be discovered as a narrow $D\bar{D}$ resonance,} up to a mass of about 
$4000\mev$.

In their study of $B^{+} \to K^{+} \psi(3770)$ decays, the Belle 
Collaboration~\cite{Abe:2003zv} has set 90\% CL upper limits on the 
transition $B^{+} \to K^{+} X(3872)$, followed by $X(3872) \to 
D\bar{D}$. Their limits imply that
\begin{eqnarray}
    \mathcal{B}(X(3872) \to D^{0}\bar{D}^{0}) & \alt & 
    4\mathcal{B}(X \to \pi^{+}\pi^{-}\jpsi)\; , \nonumber \\[-6pt]
     & & 
    \label{eq:Xddbar}  \\[-6pt]
    \mathcal{B}(X(3872) \to D^{+}D^{-})  & \alt & 
    3\mathcal{B}(X \to \pi^{+}\pi^{-}\jpsi)\; . \nonumber 
\end{eqnarray}
This constraint is already intriguingly close to the level at which we 
would expect to see $1\slj{3}{3}{3} \to D\bar{D}$.

The constraint on the total width of $X(3872)$ raises more of a 
challenge for the 2\slj{1}{2}{1} candidate, whose
$s$-wave decay to $D^{0}\bar{D}^{*0}$ rises dramatically from 
threshold, 
as shown in the middle right panel of Figure~\ref{figure:OCdecays}. 
Within the current uncertainty ($3871.7 \pm 0.6\mev$) in the mass of 
$X$, the issue cannot be settled, but the 2\slj{1}{2}{1} 
interpretation is viable only if $X$ lies below $D^{0}\bar{D}^{*0}$ 
threshold. If a light 2\slj{1}{2}{1} does turn out to be $X(3872)$, 
then its 2\slj{3}{2}{J} partners should lie nearby. In that case, 
they should be visible as relatively narrow charm-anticharm 
resonances. At $3872\mev$, we estimate  $\Gamma(2\slj{3}{2}{1}\to D\bar{D}^{*}) 
\approx 21\mev$ and $\Gamma(2\slj{3}{2}{2}\to D\bar{D}) \approx 
3\mev$. The bottom left panel in Figure~\ref{figure:OCdecays} shows 
that the 2\slj{3}{2}{2} level remains relatively narrow up to the 
opening of the $D^{*}\bar{D}^{*}$ threshold.

The 2\slj{3}{2}{0} state is an interesting special case, as 
illustrated in the bottom right panel of Figure~\ref{figure:OCdecays}. Through the
interplay of nodes in the radial wave function, 
form-factor effects, and the opening of new channels, 
$\Gamma(2\slj{3}{2}{0}\to D\bar{D})$ decreases from $\approx 60\mev$ 
at $3872\mev$ to about $12\mev$ near $3930\mev$. The total width for 
decay to open charm then rises in steps as $M(2\slj{3}{2}{0})$ 
increases through the $D_{s}\bar{D}_{s}$ and $D^{*}\bar{D}^{*}$ 
thresholds. To estimate the competing annihilation decay rate, we scale
$\Gamma(2\slj{3}{2}{0} \to gg \to \mathrm{hadrons}) \approx 
\Gamma(1\slj{3}{2}{0} \to gg \to \mathrm{hadrons}) \cdot 
\left|R_{2\mathrm{P}}^{\prime}(0)\right|^{2}/\left|R_{1\mathrm{P}}^{\prime}(0)\right|^{2}$,
where $R^{\prime}(0)$ is the derivative of the radial wave function 
at the origin. This yields $\Gamma(2\slj{3}{2}{0} \to gg \to 
\mathrm{hadrons}) \approx 1.36 \times 10.6\mev = 14.4\mev$~\cite{PDBook}, an 
estimate that should probably be reduced by the $\ket{2\slj{3}{2}{0}}$ 
fraction of the physical 2\slj{3}{2}{0} state.

We call attention to one more candidate for a narrow resonance of charmed 
mesons: The 1\slj{3}{4}{4} level remains narrow 
($\Gamma(1\slj{3}{4}{4} \to \text{charm}) \alt 
5\mev$) up to the 
$D^{*}\bar{D}^{*}$ threshold. Its allowed decays into $D\bar{D}$ and 
$D\bar{D}^{*}$ are inhibited by $\ell = 4$ barrier factors, whereas 
the $D^{*}\bar{D}^{*}$ channel is reached by $\ell = 2$.

\begin{table}
    \caption{Partial widths for decays of charmonium states into open
    charm, computed in the Cornell coupled-channel model.  All masses
    and widths are in MeV. Only significant partial widths are
    tabulated, and ``total'' refers to the sum of open-charm decays.
    Properties of the candidate states for X(3872) and their partners
    are evaluated at $3872\mev$ and also at the potential-model
    centroid for each state.  Decays occur with orbital angular
    momentum $\ell$.  For $D\bar{D}^{*}$ modes, 
    the sum of $D\bar{D}^{*}$ and $\bar{D}D^{*}$ is always implied. \label{table:OCdecays}}
    \begin{ruledtabular}
\begin{tabular}{cccccc}
    State & Mass  & $\ell$ & Channel &  Width  & 
    Total Width   \\
    \hline
     & & & & & \\[-9pt]
    1\slj{3}{3}{1} & 3770 & 1 & $\begin{array}{c} 
    D^{0}\bar{D}^{0}\\ D^{+}D^{-}\end{array}$ & 
    $\begin{array}{c} 11.8 \\ 8.3 \end{array}$ & 20.1  \\
    \hline
     & & & & & \\[-9pt]
    1\slj{3}{3}{2} & 3815 & -- & --- & 0 & 0  \\[6pt]
    1\slj{3}{3}{2} & 3872 & 1 & $D^{0}\bar{D}^{*0}$ & 0.045 & 0.045  
    \\
    \hline
     & & & & & \\[-9pt]
    1\slj{1}{3}{2} & 3815 & -- & --- & 0 & 0   \\[6pt]
    1\slj{1}{3}{2} & 3872 & 1 & $D^{0}\bar{D}^{*0}$ & 0.030 & 0.030   
    \\[6pt]
    1\slj{1}{3}{2} & 3880 & 1 & $D^{0}\bar{D}^{*0}$ & 1.7 & 1.7  \\
   \hline
     & & & & & \\[-9pt]
    1\slj{3}{3}{3} & 3872 & 3 & $\begin{array}{c} D^{0}\bar{D}^{0} \\ 
    D^{+}D^{-} \end{array}$ & $\begin{array}{c} 0.47 \\ 0.39 
    \end{array}$ & 0.86  \\[9pt]
    1\slj{3}{3}{3} & 3902 & 3 & $\begin{array}{c} D^{0}\bar{D}^{0} \\ 
    D^{+}D^{-} \end{array}$ & $\begin{array}{c} 0.84 \\ 0.72 
    \end{array}$ & 1.56  \\
    \hline
     & & & & & \\[-9pt]
    2\slj{3}{2}{0} & 3872 & 0 & $\begin{array}{c} D^{0}\bar{D}^{0} \\ 
    D^{+}D^{-} \end{array}$ & $\begin{array}{c} 27 \\ 32
    \end{array}$ & 59 \\[9pt]
    2\slj{3}{2}{0} & 3930 & 0 & $\begin{array}{c} D^{0}\bar{D}^{0} \\ 
    D^{+}D^{-} \end{array}$ & $\begin{array}{c} 5.0 \\ 7.4
    \end{array}$ & 12.4   \\[9pt]
    2\slj{3}{2}{0} & 3968 & 0 & $\begin{array}{c} D^{0}\bar{D}^{0} \\ 
    D^{+}D^{-} \\ D_{s}\bar{D}_{s} \end{array}$ & $\begin{array}{c} 
    0.27 
    \\ 0.85 \\ 40
    \end{array}$ & 41.1  \\
    \hline
     & & & & & \\[-9pt]
    2\slj{3}{2}{1} & 3872 & 0 & $D^{0}\bar{D}^{*0}$ & 20.9 & 20.9  \\[6pt]
    2\slj{3}{2}{1} & 3968 & 0 & $\begin{array}{c} D^{0}\bar{D}^{*0} \\ 
    D^{+}D^{*-} \end{array}$ & $\begin{array}{c} 71.4 \\ 78.9
    \end{array}$ & 150.3  \\
    \hline
     & & & & & \\[-9pt]
    2\slj{1}{2}{1} & 3871.6 & 0 & $D^{0}\bar{D}^{*0}$ & 4.28 & 4.28  
    \\[6pt]
    2\slj{1}{2}{1} & 3968 & 0 & $\begin{array}{c} D^{0}\bar{D}^{*0} \\ 
    D^{+}D^{*-} \end{array}$ & $\begin{array}{c} 35.5 \\ 39.2
    \end{array}$ &  74.7 \\
    \hline
     & & & & & \\[-9pt]
    2\slj{3}{2}{2} & 3872 & 2 & $\begin{array}{c} D^{0}\bar{D}^{0} \\ 
    D^{+}D^{-} \end{array}$ & $\begin{array}{c} 1.63 \\ 1.42 
    \end{array}$ & 3.05  \\[9pt]
    2\slj{3}{2}{2} & 3968 & 2 & $\begin{array}{c} D^{0}\bar{D}^{0} \\ 
    D^{+}D^{-} \\ D^{0}\bar{D}^{*0} \\ 
    D^{+}D^{*-} \end{array}$ & $\begin{array}{c} 4.4 \\ 4.2 \\ 2.57 
    \\ 2.16 
    \end{array}$ & 13.4  \\
    \hline
     & & & & & \\[-9pt]
    1\slj{3}{4}{2} & 4054 & 2 & $\begin{array}{c} D^{0}\bar{D}^{0} \\ 
    D^{+}D^{-} \\ D_{s}\bar{D}_{s} \\ D^{0}\bar{D}^{*0} \\ 
    D^{+}D^{*-} \end{array}$  & $\begin{array}{c} 46 \\ 45 \\ 4 
    \\ 31 \\ 29
    \end{array}$ & 155  \\
    \hline
     & & & & & \\[-9pt]
    1\slj{3}{4}{3} & 4054 & 2 & $\begin{array}{c} D^{0}\bar{D}^{*0} \\ 
    D^{+}D^{*-} \end{array}$ & $\begin{array}{c} 44.9 \\ 42.6 
    \end{array}$  & 87.5  \\
    \hline
     & & & & & \\[-9pt]
    1\slj{1}{4}{3} & 4054 & 2 & $\begin{array}{c} D^{0}\bar{D}^{*0} \\ 
    D^{+}D^{*-} \end{array}$ & $\begin{array}{c} 32.2 \\ 30.8
    \end{array}$ & 66.4  \\
    \hline
     & & & & & \\[-9pt]
    1\slj{3}{4}{4} & 4054 & 4 & $\begin{array}{c} D^{0}\bar{D}^{0} \\ 
    D^{+}D^{-} \\ D^{0}\bar{D}^{*0} \\ 
    D^{+}D^{*-} \end{array}$ & $\begin{array}{c} 1.96 \\ 1.78 \\ 
    0.62 \\ 0.50
    \end{array}$ & 4.88\\
\end{tabular}
\end{ruledtabular}
\end{table}

\section{For the Future \label{sec:next}}
On the experimental front, the first order of business is to establish
the nature of $X(3872)$.  Determining the spin-parity of $X$ will
winnow the field of candidates.  The charmonium interpretation and its
prominent rivals require that $X(3872)$ be a neutral isoscalar.  Are
there charged partners?  A search for $X(3872) \to \pi^{0}\pi^{0}\jpsi$
will be highly informative.  As Barnes \& Godfrey~\cite{Barnes:2003vb}
have remarked, observing a significant $\pi^{0}\pi^{0}\jpsi$ signal
establishes that $X$ is odd under charge conjugation.  Voloshin has
commented~\cite{VoloshinPC} that the ratio $\mathcal{R}_{0} \equiv
\Gamma(X\to \pi^{0}\pi^{0}\jpsi)/\Gamma(X \to \pi^{+}\pi^{-}\jpsi)$
measures the dipion isospin.  Writing $\Gamma_{I}\equiv \Gamma(X \to
(\pi^{+}\pi^{-})_{I}\jpsi)$, we see that $\mathcal{R}_{0} =
\cfrac{1}{2}/(1 + \Gamma_{1}/\Gamma_{0})$, up to kinematic corrections.
Deviations from $\mathcal{R}_{0}=\cfrac{1}{2}$ signal the
isospin-violating decay of an isoscalar, or the isospin-conserving
decay of an isovector.  Radiative decay rates and the prompt (as
opposed to $B$-decay) production fraction will provide important
guidance.  Other diagnostics of a general nature have been discussed
in Refs.~\cite{Pakvasa:2003ea,Barnes:2003vb,Close:2003sg,Close:2003mb}.

Within the charmonium framework, $X(3872)$ is most naturally 
interpreted as the 1\slj{3}{3}{2} or 1\slj{3}{3}{3} level, both of 
which have allowed decays into $\pi\pi\jpsi$. The 
$2^{--}$ 1\slj{3}{3}{2} state is forbidden by parity conservation to 
decay into $D\bar{D}$ but has a modest $D^{0}\bar{D}^{*0}$ partial width 
for masses near $3872\mev$. Although the uncertain $\pi\pi\jpsi$ 
partial width makes it difficult to estimate relative branching 
ratios, the decay $X(3872) \to \chi_{c1}\,\gamma(344)$ should show itself 
if $X$ is indeed 1\slj{3}{3}{2}. The $\chi_{c2}\,\gamma(303)$ line 
should be seen with about $\cfrac{1}{4}$ the strength of 
$\chi_{c1}\,\gamma(344)$.  In our coupled-channel calculation, the 
1\slj{3}{3}{2} mass is about $41\mev$ lower than the observed 
$3872\mev$. In contrast, the computed 1\slj{3}{3}{3} mass is quite 
close to $3872\mev$, and 1\slj{3}{3}{3} does not have an E1 
transition to $\chi_{c1}\,\gamma(344)$. The dominant decay of the 
$3^{--}$ 1\slj{3}{3}{3} state should be into $D\bar{D}$; a small 
branching fraction for the $\pi\pi\jpsi$ discovery 
mode would imply a large production rate. One radiative 
transition should be observable, with $\Gamma(X(3872) \to 
\chi_{c2}\,\gamma(303)) \agt \Gamma(X(3872) \to \pi^{+}\pi^{-}\jpsi)$. 
We underscore the importance of searching for the $\chi_{c1}\,\gamma(344)$ 
and $\chi_{c2}\,\gamma(303)$ lines.
    
Beyond pinning down the character of $X(3872)$, experiments can search 
for additional narrow charmonium states in radiative and hadronic 
transitions to lower-lying $c\bar{c}$ levels, as we emphasized in 
Ref.~\cite{Eichten:2002qv}, and in neutral combinations of charmed 
mesons and anticharmed mesons. The coupled-channel analysis presented 
in this paper sets up specific targets.

On the theoretical front, we need a more complete understanding of the
production of the charmonium states in $B$ decays and by direct
hadronic production, including the influence of open-charm channels.
Understanding of the production mechanisms for molecular charm or
$c\bar{c}g$ hybrid states is much more primitive.  We need to improve
the theoretical understanding of hadronic cascades among charmonium
states, including the influence of open-charm channels.  The comparison
of charmonium transitions with their upsilon counterparts should be
informative.  The analysis we have carried out can be extended to the
$b\bar{b}$ system, where it may be possible to see discrete
threshold-region states in direct hadronic production.  Because the
Cornell coupled channel model is only an approximation to QCD, it would
be highly desirable to compare its predictions with those of a
coupled-channel analysis of the \slj{3}{2}{0} model of quark pair
production. Ultimately, extending lattice QCD calculations into the 
flavor-threshold region should give a firmer basis for predictions.

In addition to the $1\slj{1}{2}{1}\;h_{c}$, the now-established 
$2\slj{1}{1}{0}\;\eta_{c}^{\prime}$, and the long-sought 
$1\slj{1}{3}{2}\;\eta_{c2}$ and $1\slj{3}{3}{2}\;\psi_{2}$ states, 
discrete charmonium levels are to be found as narrow charm-anticharm 
structures in the flavor-threshold region. The most likely candidates 
correspond to the $1\slj{3}{3}{3}$, $2\slj{3}{2}{2}$, and 
$1\slj{3}{4}{4}$ levels. If $X(3872)$ is indeed a charmonium 
state---the \slj{3}{3}{2} and \slj{3}{3}{3} assignments seem most 
promising---then identifying that state anchors the mass scale. If 
$X(3872)$ is not charmonium, then all the charmonium levels remain to 
be discovered.
Finding these states---and establishing their 
masses, widths, and production rates---will lead us into new terrain.

\begin{acknowledgments}
We thank Steve Olsen, J. D. Jackson, Stephen Godfrey, Gerry Bauer, Brad 
Abbott, Vivek Jain, Sheldon Stone, and 
participants in the Quarkonium Working Group's Second Workshop on Heavy 
Quarkonium for inspiring discussions.   KL's research was supported by the Department of Energy
under Grant~No.~DE--FG02--91ER40676.  Fermilab is operated by
Universities Research Association Inc.\ under Contract No.\
DE-AC02-76CH03000 with the U.S.\ Department of Energy.


\end{acknowledgments}
\bibliography{Diagnostics}

\begin{thebibliography}{22}
\expandafter\ifx\csname natexlab\endcsname\relax\def\natexlab#1{#1}\fi
\expandafter\ifx\csname bibnamefont\endcsname\relax
  \def\bibnamefont#1{#1}\fi
\expandafter\ifx\csname bibfnamefont\endcsname\relax
  \def\bibfnamefont#1{#1}\fi
\expandafter\ifx\csname citenamefont\endcsname\relax
  \def\citenamefont#1{#1}\fi
\expandafter\ifx\csname url\endcsname\relax
  \def\url#1{\texttt{#1}}\fi
\expandafter\ifx\csname urlprefix\endcsname\relax\def\urlprefix{URL }\fi
\providecommand{\bibinfo}[2]{#2}
\providecommand{\eprint}[2][]{\url{#2}}

\bibitem[{\citenamefont{Choi et~al.}(2002)}]{Choi:2002na}
\bibinfo{author}{\bibfnamefont{S.~K.} \bibnamefont{Choi}} \bibnamefont{et~al.}
  (\bibinfo{collaboration}{Belle}), \bibinfo{journal}{Phys. Rev. Lett.}
  \textbf{\bibinfo{volume}{89}}, \bibinfo{pages}{102001}
  (\bibinfo{year}{2002}), \bibinfo{note}{erratum: ibid.\ 89, 129901 (2002)},
  \eprint{hep-ex/0206002}.

\bibitem[{\citenamefont{Eichten et~al.}(2002)\citenamefont{Eichten, Lane, and
  Quigg}}]{Eichten:2002qv}
\bibinfo{author}{\bibfnamefont{E.~J.} \bibnamefont{Eichten}},
  \bibinfo{author}{\bibfnamefont{K.}~\bibnamefont{Lane}}, \bibnamefont{and}
  \bibinfo{author}{\bibfnamefont{C.}~\bibnamefont{Quigg}},
  \bibinfo{journal}{Phys. Rev. Lett.} \textbf{\bibinfo{volume}{89}},
  \bibinfo{pages}{162002} (\bibinfo{year}{2002}), \eprint{hep-ph/0206018}.

\bibitem[{\citenamefont{Choi et~al.}(2003)}]{Choi:2003ue}
\bibinfo{author}{\bibfnamefont{S.~K.} \bibnamefont{Choi}} \bibnamefont{et~al.}
  (\bibinfo{collaboration}{Belle}), \bibinfo{journal}{Phys. Rev. Lett.}
  \textbf{\bibinfo{volume}{91}}, \bibinfo{pages}{262001}
  (\bibinfo{year}{2003}), \eprint{hep-ex/0309032}.

\bibitem[{\citenamefont{Acosta et~al.}(2003)}]{Acosta:2003zx}
\bibinfo{author}{\bibfnamefont{D.}~\bibnamefont{Acosta}} \bibnamefont{et~al.}
  (\bibinfo{collaboration}{CDF II Collaboration}) (\bibinfo{year}{2003}),
  \eprint{hep-ex/0312021}.

\bibitem[{\citenamefont{D\O~Collaboration}(2003)}]{Dzerotemp}
\bibinfo{author}{\bibfnamefont{D\O~Collaboration}}
(\bibinfo{year}{2004}),
  \bibinfo{note}{{{D\O} Note 4334}}, \urlprefix\url{http://www-d0.fnal.gov
  /Run2Physics/ckm/Moriond_2003/X_conf_note_v9.ps}.
  
\bibitem[{\citenamefont{Asner}(2003)}]{Asner:2003wv}
\bibinfo{author}{\bibfnamefont{D.~M.} \bibnamefont{Asner}}
  (\bibinfo{collaboration}{CLEO}) (\bibinfo{year}{2003}),
  \eprint{hep-ex/0312058}.

\bibitem[{\citenamefont{Wagner}(2003)}]{Wagner:2003qb}
\bibinfo{author}{\bibfnamefont{G.}~\bibnamefont{Wagner}}
  (\bibinfo{collaboration}{BABAR}) (\bibinfo{year}{2003}),
  \eprint{hep-ex/0305083}.

\bibitem[{\citenamefont{Abe et~al.}(2003{\natexlab{a}})}]{Abe:2003ja}
\bibinfo{author}{\bibfnamefont{K.}~\bibnamefont{Abe}} \bibnamefont{et~al.}
  (\bibinfo{collaboration}{Belle}) (\bibinfo{year}{2003}{\natexlab{a}}),
  \eprint{hep-ex/0306015}.

\bibitem[{\citenamefont{Skwarnicki}(2003)}]{Skwarnicki:2003wn}
\bibinfo{author}{\bibfnamefont{T.}~\bibnamefont{Skwarnicki}}
  (\bibinfo{year}{2003}), \eprint{hep-ph/0311243}.

\bibitem[{\citenamefont{Eichten et~al.}(1978)\citenamefont{Eichten, Gottfried,
  Kinoshita, Lane, and Yan}}]{Eichten:1978tg}
\bibinfo{author}{\bibfnamefont{E.}~\bibnamefont{Eichten}},
  \bibinfo{author}{\bibfnamefont{K.}~\bibnamefont{Gottfried}},
  \bibinfo{author}{\bibfnamefont{T.}~\bibnamefont{Kinoshita}},
  \bibinfo{author}{\bibfnamefont{K.~D.} \bibnamefont{Lane}}, \bibnamefont{and}
  \bibinfo{author}{\bibfnamefont{T.-M.} \bibnamefont{Yan}},
  \bibinfo{journal}{Phys. Rev.} \textbf{\bibinfo{volume}{D17}},
  \bibinfo{pages}{3090} (\bibinfo{year}{1978}), \bibinfo{note}{[Erratum-ibid.\
  D {\bf 21}, 313 (1980)].}

\bibitem[{\citenamefont{Eichten et~al.}(1980)\citenamefont{Eichten, Gottfried,
  Kinoshita, Lane, and Yan}}]{Eichten:1980ms}
\bibinfo{author}{\bibfnamefont{E.}~\bibnamefont{Eichten}},
  \bibinfo{author}{\bibfnamefont{K.}~\bibnamefont{Gottfried}},
  \bibinfo{author}{\bibfnamefont{T.}~\bibnamefont{Kinoshita}},
  \bibinfo{author}{\bibfnamefont{K.~D.} \bibnamefont{Lane}}, \bibnamefont{and}
  \bibinfo{author}{\bibfnamefont{T.-M.} \bibnamefont{Yan}},
  \bibinfo{journal}{Phys. Rev.} \textbf{\bibinfo{volume}{D21}},
  \bibinfo{pages}{203} (\bibinfo{year}{1980}).

\bibitem[{\citenamefont{Barnes and Godfrey}(2003)}]{Barnes:2003vb}
\bibinfo{author}{\bibfnamefont{T.}~\bibnamefont{Barnes}} \bibnamefont{and}
  \bibinfo{author}{\bibfnamefont{S.}~\bibnamefont{Godfrey}}
  (\bibinfo{year}{2003}), \eprint{hep-ph/0311162}.

\bibitem[{\citenamefont{{Hagiwara} et~al.}(2002)}]{PDBook}
\bibinfo{author}{\bibfnamefont{K.}~\bibnamefont{{Hagiwara}}}
  \bibnamefont{et~al.} (\bibinfo{collaboration}{Particle Data Group}),
  \bibinfo{journal}{{Phys. Rev. D}} \textbf{\bibinfo{volume}{66}},
  \bibinfo{pages}{010001} (\bibinfo{year}{2002}), \bibinfo{note}{and 2003
  off-year partial update}, \urlprefix\url{http://pdg.lbl.gov}.

\bibitem[{\citenamefont{Pakvasa and Suzuki}(2004)}]{Pakvasa:2003ea}
\bibinfo{author}{\bibfnamefont{S.}~\bibnamefont{Pakvasa}} \bibnamefont{and}
  \bibinfo{author}{\bibfnamefont{M.}~\bibnamefont{Suzuki}},
  \bibinfo{journal}{Phys. Lett.} \textbf{\bibinfo{volume}{B579}},
  \bibinfo{pages}{67} (\bibinfo{year}{2004}), \eprint{hep-ph/0309294}.

\bibitem[{\citenamefont{T\"{o}rnqvist}(2004)}]{Tornqvist:2004qv}
\bibinfo{author}{\bibfnamefont{N.~A.} \bibnamefont{T\"{o}rnqvist}}
(\bibinfo{year}{2004}),  \eprint{hep-ph/0402237}.

\bibitem[{\citenamefont{Bai et~al.}(2003)}]{Bai:2003hv}
\bibinfo{author}{\bibfnamefont{J.~Z.} \bibnamefont{Bai}} \bibnamefont{et~al.}
  (\bibinfo{collaboration}{BES}) (\bibinfo{year}{2003}),
  \eprint{hep-ex/0307028}.

\bibitem[{\citenamefont{Eichten and Quigg}(1994)}]{Eichten:1994gt}
\bibinfo{author}{\bibfnamefont{E.~J.} \bibnamefont{Eichten}} \bibnamefont{and}
  \bibinfo{author}{\bibfnamefont{C.}~\bibnamefont{Quigg}},
  \bibinfo{journal}{Phys. Rev.} \textbf{\bibinfo{volume}{D49}},
  \bibinfo{pages}{5845} (\bibinfo{year}{1994}).

\bibitem[{\citenamefont{Le~Yaouanc et~al.}(1977)\citenamefont{Le~Yaouanc,
  Oliver, P\`{e}ne, and Raynal}}]{LeYaouanc:1977ux}
\bibinfo{author}{\bibfnamefont{A.}~\bibnamefont{Le~Yaouanc}},
  \bibinfo{author}{\bibfnamefont{L.}~\bibnamefont{Oliver}},
  \bibinfo{author}{\bibfnamefont{O.}~\bibnamefont{P\`{e}ne}}, \bibnamefont{and}
  \bibinfo{author}{\bibfnamefont{J.~C.} \bibnamefont{Raynal}},
  \bibinfo{journal}{Phys. Lett.} \textbf{\bibinfo{volume}{B71}},
  \bibinfo{pages}{397} (\bibinfo{year}{1977}).

\bibitem[{\citenamefont{Abe et~al.}(2003{\natexlab{b}})}]{Abe:2003zv}
\bibinfo{author}{\bibfnamefont{K.}~\bibnamefont{Abe}} \bibnamefont{et~al.}
  (\bibinfo{collaboration}{Belle}) (\bibinfo{year}{2003}{\natexlab{b}}),
  \eprint{hep-ex/0307061}.

\bibitem[{\citenamefont{Voloshin}(2003)}]{VoloshinPC}
\bibinfo{author}{\bibfnamefont{M.}~\bibnamefont{Voloshin}}
  (\bibinfo{year}{2003}), \bibinfo{note}{private communication}.

\bibitem[{\citenamefont{Close and Page}(2004)}]{Close:2003sg}
\bibinfo{author}{\bibfnamefont{F.~E.} \bibnamefont{Close}} \bibnamefont{and}
  \bibinfo{author}{\bibfnamefont{P.~R.} \bibnamefont{Page}},
  \bibinfo{journal}{Phys. Lett.} \textbf{\bibinfo{volume}{B578}},
  \bibinfo{pages}{119} (\bibinfo{year}{2004}), \eprint{hep-ph/0309253}.

\bibitem[{\citenamefont{Close and Godfrey}(2003)}]{Close:2003mb}
\bibinfo{author}{\bibfnamefont{F.~E.} \bibnamefont{Close}} \bibnamefont{and}
  \bibinfo{author}{\bibfnamefont{S.}~\bibnamefont{Godfrey}},
  \bibinfo{journal}{Phys. Lett.} \textbf{\bibinfo{volume}{B574}},
  \bibinfo{pages}{210} (\bibinfo{year}{2003}), \eprint{hep-ph/0305285}.

\end{thebibliography}

\end{document}